\documentclass[aps,physrev,preprint,superscriptaddress]{revtex4-2}

\usepackage{graphicx}
\usepackage{hyperref}
\usepackage{amsmath,amssymb,xcolor}
\usepackage{upgreek}

\begin{document}


\title{Ridged geometries induce axial flow vortices in Couette systems}


\author{Akankshya Majhi}
\affiliation{Physical Chemistry and Soft Matter, Wageningen University and Research, Wageningen, The Netherlands}

\author{Maria Raquel Serial}
\affiliation{Laboratory of Biophysics, \\ Wageningen University and Research, Wageningen, The Netherlands}
\affiliation{Department of Process and Energy, Delft University of Technology, Delft, The Netherlands}

\author{Lars Kool}
\affiliation{Physical Chemistry and Soft Matter, Wageningen University and Research, Wageningen, The Netherlands}
\affiliation{Laboratoire de Physique et M\'echanique des Milieux H\'et\'erog\`enes (PMMH), UMR 7636 CNRS, ESPCI Paris, PSL Research University, Sorbonne Universit\'e and Universit\'e Paris Cit\'e,
7-9 Quai Saint-Bernard, Paris, 75005, France}
\affiliation{Institut Pierre Gilles de Gennes (IPGG), ESPCI, Paris, France }

\author{Jasper~van der Gucht}
\affiliation{Physical Chemistry and Soft Matter, Wageningen University and Research, Wageningen, The Netherlands}

\author{Joshua A. Dijksman}
\email{j.a.dijksman@uva.nl}
\affiliation{Van der Waals-Zeeman Institute, Institute of Physics, University of Amsterdam, Amsterdam, The Netherlands}



\begin{abstract}
A yield stress fluid has a critical stress above which the material starts to flow. Typically, the yield stress behaviour is captured in the Herschel-Bulkley (HB) model, which assumes a constant yield stress as material parameter. It is not clear whether the simultaneous superposition of a flow in an orthogonal direction to the main flow, that displays HB behaviour, affects the yield stress and will make the yield stress either flow rate- or field-dependent. Therefore, it is important to understand how the presence of flow in two orthogonal directions affects the yielding behaviour of the fluid in general. In this work, we showed that wall patterning can be used to generate flow in two orthogonal directions simultaneously. We find that these orthogonal flows measurably affected each other. We induced spatially varying secondary flows by shearing a standard Newtonian fluid and two common yield stress fluids in a rheometer using a concentric cylinder geometry with angled ridges. We measured the normal force as a function of rotation rates for different angled ridges from conventional rheological measurements. We also imaged the flow fields by employing rheo-MRI, to directly measure the penetration depth of the fluids into the rough boundary of the ridged geometry. Finally, we relate the penetration depth to the axial flow for different geometries, at different imposed rotation rates and the fluid type, to show that two flow directions in the yield stress fluids are indeed significantly related to each other.
\end{abstract}


\maketitle


\section{Introduction}
\label{sec:intro}

Many everyday materials fall into the category of complex fluids, which are neither purely elastic solids nor simple Newtonian fluids. Yield stress fluids are an important class of such materials, and their flow behaviour remains an active area of investigation.\cite{Balmforth2014, Coussot2014, Bonn2017, Malkin2017, Accetta2024, Divoux2024} These fluids have a critical stress threshold, known as the yield stress, above which the material starts to flow. The transition from solid-like to liquid-like state is commonly referred to as yielding.\cite{Barnes1999} Examples of yield stress materials include food products, such as mayonnaise, sauces and spreads, concrete, pharmaceutical creams, and cosmetic pastes. Controlling the flow behaviour of these products is crucial for their commercial success.\cite{Meeker2004, Paredes2011, Coussot2002a} Despite their importance, yield stress fluids are known for their complex non-trivial rheological behaviour. Typically, the yield stress behaviour is well captured in the Herschel-Bulkley model: $\sigma = \sigma_y + k{\dot{\gamma}}^n$, where, $\sigma_y$ denotes the yield stress, $k$ is the consistency index and $n$ is the flow index.\cite{Herschel1926, Macosko1994} This model effectively describes yield stress fluids under simple one-directional flows. However, numerous practical applications involve more complex flow scenarios such as simultaneous multi-directional flows\cite{Shaukat2012} or flow past complex surfaces, such as corrugated plates, helical mixers, and impellers to name a few. The behaviour of yield stress fluids in such spatially varying and multi-directional flow scenarios is not fully understood. 

Yield stress fluids subjected to shear deformation are known to develop normal stresses, as extensively studied in classical works.\cite{Bird1987, Macosko1994, Larson1999} Normal stress differences arise from microscopic mechanisms, which vary depending on the fluid type. For example, in emulsions, deformation of suspended droplets under shear is believed to produce finite normal stress differences, analogous to the behaviour observed in polymers.\cite{deCagny2019} However, the sign and physical origins of these stresses\textemdash whether positive or negative\textemdash remain topics of ongoing debate\cite{Gauthier2021} and the previous literature on normal stress measurements for these materials are scarce. The few existing studies do not offer a coherent understanding, partly due to the experimental challenges posed by flow heterogeneities.\cite{deCagny2019} Notably, yield stress fluids exhibit a normal yield stress, wherein normal stresses do not vanish as the shear rate approaches zero,\cite{deCagny2019} highlighting the intrinsic relationship between shear and normal stress components in these materials. These complexities emphasise the need for further experimental and theoretical efforts to better understand and quantify normal stresses in yield stress fluids, particularly under heterogeneous flow conditions.\cite{Habibi2016, Montesi2004}

There are studies on simultaneous multi-directional flows of various complex fluids. For example, Ovarlez \textit{et al.}\cite{Ovarlez2010} imposed a squeeze flow at a constant squeeze shear rate while simultaneously providing a rotational shear rate ramp to the material. They showed that in the case of emulsions, when a flow is imposed in one direction, the secondary flow has no yield resistance (i.e., no yield stress), which implies that the system is always unjammed  simultaneously in all flow directions. Additionally, Lin and coworkers\cite{Lin2016}
have demonstrated that the shear thickening behaviour in suspensions can be tuned by manipulating flows imposed in two orthogonal directions, in particular by superimposing a small orthogonal perturbation in addition to the primary shear direction. 

\subsubsection*{Inducing and imaging orthogonal flow}

\begin{figure}[h!]
\centering
\includegraphics[scale=0.5]{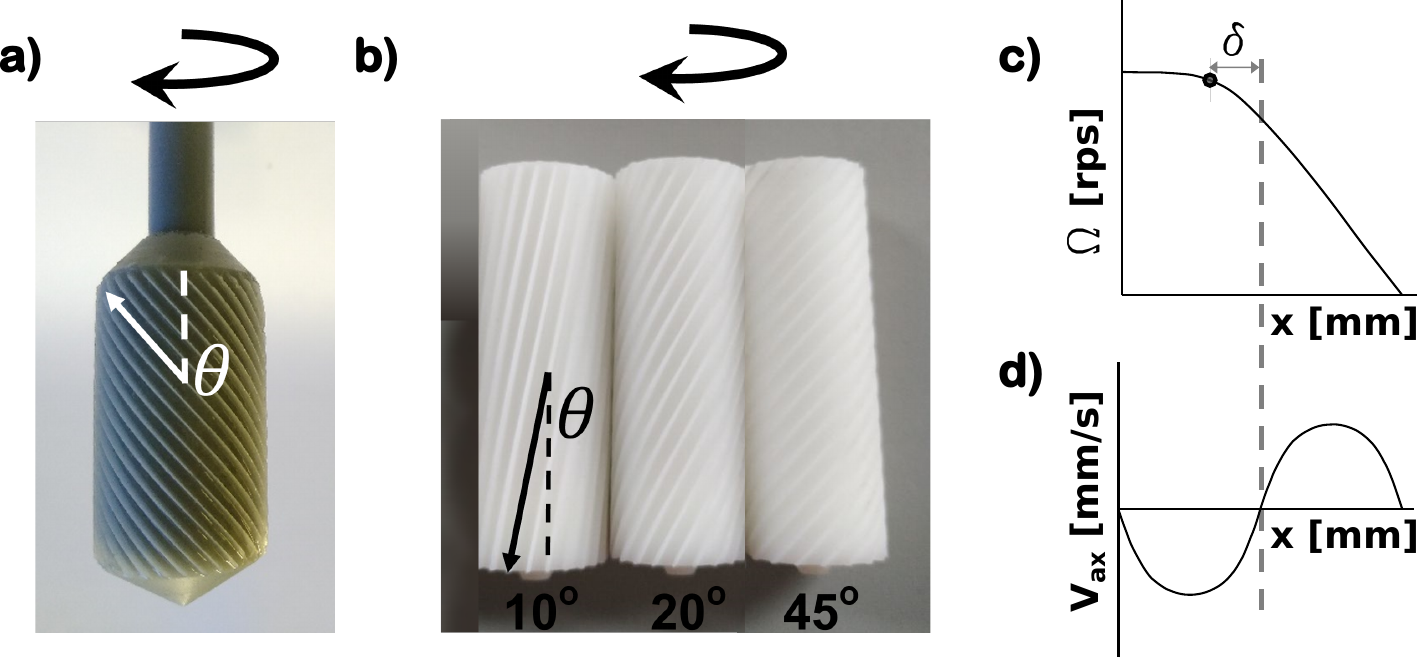}
\caption{\textbf{a)} Angled ridged geometry used in the rheological experiments, indicating the ridges aligned at an angle $\theta$ to the vertical axis. \textbf{b)} Angled ridged geometries used in the rheo-MRI experiments, with ridges at an angle $\theta$ relative to the vertical axis. Curved arrows in \textbf{a} and \textbf{b} indicate the direction of rotation. Note the different orientations of the ridges in the two geometries in \textbf{a} and \textbf{b}.  \textbf{c)} Schematic representation of the fluid rotation rate (in the azimuthal direction) as a function of position in the gap. The grey dashed line indicates the tip of the ridge. Between the ridges, fluid moves with the geometry, while Taylor-Couette flow is observed outside the ridges. $\delta$ is the penetration depth, that we measure from this profile. \textbf{d)}~Schematic representation of the mean axial velocity as a function of position in the gap. The grey dashed line indicates the tip of the ridge, with fluid being pushed downward (or upward if the rotation direction is reversed) between the ridges, and the fluid  outside the ridges moving upward to maintain the overall volume. Negative axial velocities point in the direction of gravity.}
\label{fig:rheo_rheoMRI_geo}
\end{figure}

These previous studies highlight the importance of understanding the behaviour of yield stress fluids in multi-directional flows. This work primarily aims to investigate the relationship between flows in two orthogonal directions, by utilising wall patterning to generate such flows. The secondary flow is induced via shearing a standard Newtonian fluid and two common yield stress fluids in a rheometer using a concentric cylinder geometry with patterned walls (Figure~\ref{fig:rheo_rheoMRI_geo}). These geometries were 3D printed and were used in our previous work\cite{Majhi2024} to study the effect of wall boundaries on the flow behaviour of a simple Newtonian fluid. We showed that in such patterned geometries, the flow profiles penetrate up to a certain extent in between the ridges and we denote the extent of this penetration with a parameter, called the penetration depth $\delta$, consistent with our earlier work.\cite{Majhi2024} In case of geometries with ridges at an angle, we found that the fluid penetration automatically leads to a secondary axial flow in addition to the primary imposed flow direction. It is to be noted that the secondary flow in our work varies spatially due to the helical pattern of the geometry. Our geometries are different from those used in the previous combined rotational shear and squeeze flow studies by Ovarlez\cite{Ovarlez2010} in the sense that they are conceptually simple and give access to steady state shear situations at a constant gap. Secondly, such geometries are useful to measure yield stress in slow flow situations. Additionally, we could easily measure normal force and torque, and subsequently two stress components, for a wide range of imposed rotation rates \textemdash a capability not achievable with the setup employed by Ovarlez. 

In addition to the traditional rheological measurements, we measure the local flow behaviour of these test fluids through magnetic resonance imaging (MRI) technique.\cite{Callaghan1999, Callaghan2011, Hollingsworth2004, Serial2019} These local measurements are useful to study local effects such as wall slip and shear localisation etc.\cite{Bonn2017} In this current work, we aim to extend our understanding of the penetration depth in non-Newtonian fluids by directly measuring it from the rheo-MRI velocity profiles for different imposed rotation rates and ridge angles. This knowledge will provide deeper insights on how complex fluids behave in spatially varying flow situations.          

The current paper is organised as follows. First, we present the preparation methodology of the test fluids, the ridged geometries and the experimental setups. Second, we report the corresponding results from the rheological and rheo-MRI experiments. Finally, we compare the two local velocity profiles by extracting two relevant parameters from them and then discuss how these two parameters affect the overall bulk flow behaviour.

\section{Experimental details}
\label{sec:expdetails}

To visualise the fluid flow penetration in rough geometries, we examined three test fluids: a Newtonian fluid and two yield stress fluids. Castor oil (Sigma Aldrich, CAS no.\ 8001-79-4) was selected as the Newtonian fluid due to its constant viscosity of $\sim$1 Pa s at 20~$^\circ$C. For the yield stress fluids, we used a model system and a dense emulsion. The preparation methods for these yield stress fluids are detailed in the following subsection.

\subsection{Yield stress fluids}

In this work, we used two yield stress fluids: a Carbopol suspension and a dense emulsion. Carbopol is widely used as a model yield stress fluid\cite{Piau2007, Moller2009a, Ovarlez2013} that exhibits little to no thixotropy.\cite{Coussot2009, Moller2009b, Coussot2014, Dinkgreve2017, deCagny2019} It is composed of particles made of a crosslinked polyacrylic acid polymer, typically provided as a dry white powder, and forms a yield stress fluid when dissolved in water due to the swelling of polymer coils.\cite{Dinkgreve2018} Despite its classification as a polymeric gel, all interparticle interactions in Carbopol are repulsive, allowing it to be considered a soft particle emulsion rather than a traditional gel.\cite{Vlassopoulos2014} Additionally, Carbopol is stable over time and does not require extensive solvent or density and concentration matching, making it advantageous compared to other jammed systems. Its simple preparation and easy cleaning from 3D printed geometries make it particularly suitable for our experimental setup.

The Carbopol suspension was prepared by mixing 0.5 wt\% of 1.25 MDa polyacrylic acid (Sigma Aldrich, CAS no.\ 9003-01-4) and deionised water. The solution was stirred at 500 rpm for about 2 h, until the Carbopol was completely dissolved and no visible lumps remained. Care was taken to avoid elevated temperatures, high stirring speeds, and prolonged mixing times to prevent degradation of the polymer.\cite{Dinkgreve2018} Subsequently, the pH of the solution was adjusted to 7 through dropwise addition of 1M NaOH solution. During this continuous neutralisation process, the solution gelled, requiring manual stirring to maintain homogeneity.  The prepared suspension was then left to rest at room temperature for a minimum of 12 h to allow for the dissipation of air bubbles.

In addition to Carbopol, we prepared a dense oil-in-water emulsion using castor oil, stabilized by sodium dodecyl sulphate (SDS). The oil volume fraction was 0.8 and the surfactant concentration in the aqueous phase was 1 wt$\%$. The emulsion was prepared following a protocol adapted from the works of Dekker~\textit{et~al.}\cite{Dekker2018}  and Paredes~\textit{et~al.}\cite{Paredes2013} The microstructure of our emulsion was found to be consistent with that of the emulsions made in previous studies.\cite{Dekker2018, Paredes2013} The oil droplets have a mean diameter of 3.2~$\upmu$m with a polydispersity of 20$\%$, as we determined using confocal laser scanning microscopy (refer to inset image in Appendix \ref{subsec:flow_curve_YSF}).

For both the yield stress fluids, we determined the yield stress from the steady shear measurements using standard geometries in a rheometer. The yield stresses of Carbopol and castor oil-in-water emulsion were found to be $\sim$8 Pa and $\sim$68 Pa, respectively (refer to Appendix \ref{subsec:flow_curve_YSF}).

\subsection{Design and fabrication of the ridged geometries}

For the rheological measurements, custom 3D printed ridged geometries were designed and fabricated as described in our previous work.\cite{Majhi2024} These geometries, as illustrated in Figure~\ref{fig:rheo_rheoMRI_geo}a, were all designed with the shortest distance (spacing) of 2 mm between the ridges, a ridge depth of 1 mm and a ridge width of 0.3 mm (refer to Appendix Figure~\ref{fig:rheo_rheoMRI_geo_SI}a for detailed dimensions). For the rheological measurements, we used four different geometries with ridge angles 10$^\circ$, 20$^\circ$,  40$^\circ$ and 50$^\circ$, respectively. In all the geometries, the ridge depth, width and the spacing between the ridges were kept constant. 

For the rheo-MRI measurements, we adapted the ridged geometries based on the standard CC20 geometry previously used in this rheo-MRI setup.\cite{Nikolaeva2019, Nikolaeva2020, Serial2022} The ridge depth, spacing between the ridges and ridge width were kept the same as those used in the geometries for rheological measurements. Three geometries with ridges angled at 10$^\circ$, 20$^\circ$ and 45$^\circ$ were fabricated via 3D printing. The geometries were designed to be hollow and open on one end to accommodate a reference fluid (water, in our case). To minimize wobbling caused by the long driving shaft of the rheo-MRI setup, a centering pin was printed at the bottom. Additionally, we designed a cap with a shaft to securely connect the geometry to the driving shaft. Detailed drawings and dimensions of the different parts of the geometry are given in Appendix Figures~\ref{fig:rheo_rheoMRI_geo_SI}b--e. The rheo-MRI geometries were printed and cured following the procedure outlined in our previous work,\cite{Majhi2024} with minor
modifications. The supports were placed only on the inside, and both the geometry and the cap were milled on a lathe after printing to ensure a wobble-free alignment. The cap was secured using multiple wrappings of thin PTFE tape, approximately 100 $\upmu$m thick.

\subsection{Rheological experiments}

Rheological experiments were performed on an Anton Paar MCR 501 rheometer, using the custom geometries in standard cups and cup holders (Anton Paar CC17). All the experiments were performed at 20.5 $^\circ$C. The printed geometries were tested on two test fluids: castor oil and Carbopol. Prior to each measurement, the geometry was placed in the test fluid to remove air bubbles between the ridges. The geometries were then visually inspected to ensure no air bubbles remained; any detected bubbles were eliminated by rubbing the geometry against the cup while it was submerged in the fluid.

We performed steady shear measurements with rotation rates ranging from $\Omega_\textnormal{o}$ =~10$^{-5}$ to 10$^{0}$ rps, which correspond to the laminar regime. Each measurement consisted of 26 datapoints, distributed logarithmically, and each datapoint started with the stepwise increase of $\Omega_\textnormal{o}$, followed by an equilibration period of 2~min, after which data was averaged for 5 s. Before each measurement, the test fluids were pre-sheared at a shear rate $\dot{\gamma}$ = 1000 s$^{-1}$ for 60~s to minimise the impact of stress induced by the loading of the test fluid into the measurement geometry. This was followed by a  recovery period of 60~s to allow the stress to relax\cite{deCagny2019} and to establish controlled initial conditions.\cite{Paredes2011, Coussot2002b, Barnes1997}

\subsection{Rheo-MRI experiments}

Rheo-MRI experiments were conducted on a Bruker Avance III spectrometer operating at 7 T magnetic field strength, with a resonance frequency of 300~MHz for ${^1}$H. The excitation and detection of ${^1}$H signal were performed using a birdcage radiofrequency coil with an inner diameter of 25 mm, combined with a 3D microimaging gradient system, Micro 2.5 (Bruker Biospin), with a maximum gradient intensity of 1.5 T~m$^{-1}$ along all three axes. The magnet was equipped with a standard Bruker rheo-MRI accessory,  consisting of a stepper motor and drive shaft, in addition to a serrated Couette cell (CC) geometry, with inner and outer cylinder diameters of 20~mm and 22~mm, respectively, for three different ridge angles (10$^\circ$, 20$^\circ$ and 45$^\circ$). Hence, the gap size is 1 mm. 

For the measurement of 1D velocity profiles, a Pulsed Gradient Spin Echo (PGSE) sequence\cite{Callaghan1999, Callaghan2011} was implemented with a slice selection of 1 mm $\times$ 1 mm.\cite{Nikolaeva2018, Serial2021b, Nikolaeva2020, Serial2022} Local averaged velocities along the $z$ and $y$-directions\cite{Serial2019} were measured with a field of view of 25 mm and 512 points were collected, resulting in a spatial resolution of 48.8~$\upmu$m. For all measurements, the signal to noise ratio was increased by averaging the signal from 32 or 128 excitations, depending on the noise level of the acquired profiles. 

All measurements were performed in triplicate, using the same sample for each rotation rate measurement with a particular geometry. The sample was only replaced when switching to a different angled geometry.

\subsection{Determination of penetration depth and maximum axial velocity}
\label{subsec:calcmethod_delta_maxVax}

We compute the penetration depth as the distance between the position where the rotational velocity starts to drop from its value at the rotating cylinder and the tip of the ridges ($x = 10$ mm). 

To determine the position where the rotational velocity starts to decrease from its initial plateau, we first fit the rotational  velocities with the following empirical sigmoidal function:
\begin{equation}
\Omega = \Omega_\mathrm{min} + \frac{\Delta \Omega}{1 + e^{({x_\textnormal{c} \, - \, x})k}}
\label{ch3:eq:sigmoidalfunction}
\end{equation}
\noindent where $\Delta \Omega$ is the difference between the plateau $\Omega$ (i.e., $\Omega_\mathrm{\max}$) and $\Omega_\mathrm{\min}$, $x_\textnormal{c}$ is the position at which $\Omega = \Omega_\mathrm{min} + \Delta \Omega/2$, and $k$ is related to how steep the curve is around $x = x_\textnormal{c}$. 

Using the fitted parameters $x_\textnormal{c}$ and $k$, we calculate $x_\textnormal{d}$, which provides a quantitative measure of the \textsf{`}drop-off\textsf{'} position. We define this \textsf{`}drop-off\textsf{'} position as the position in the gap where the rotational velocity decreases by 5$\%$ of the amplitude (the difference between the plateau and minimum values of the velocity profile) from its plateau value. The drop-off positions for all the fluids are marked with black circles in panels a--c of Figures~\ref{fig:velprof_Newtonian_SI}, \ref{fig:velprof_Carbopol_SI} and \ref{fig:velprof_Emulsion_SI}, respectively.

To obtain the maximum axial velocity, we identify the minimum of the axial velocity in the gap between the inner cylinder and the ridge tip, and take its absolute value. Throughout this paper, we refer to this quantity as the maximum axial velocity, $V_\textnormal{ax,max}$.

\section{Results and discussion}
\label{sec:RnD}

We investigate the flow of a Newtonian and two yield stress fluids in angled ridged geometries using conventional bulk rheological measurements and rheo-MRI measurements. First, we discuss the steady shear measurements of castor oil and Carbopol in different angled geometries. Then, we present the velocity profiles of the three test fluids for a range of imposed rotation rates in three different angled geometries. Lastly, we quantify the extent of fluid flow profile penetration between the ridges for the three different fluids and finally, we discuss the dependence of axial flow on the different geometries, different imposed rotation rates and the fluid type. 

\subsection{Bulk response from rheological experiments}

\begin{figure}[htbp!]
\centering
\includegraphics[scale=0.45]{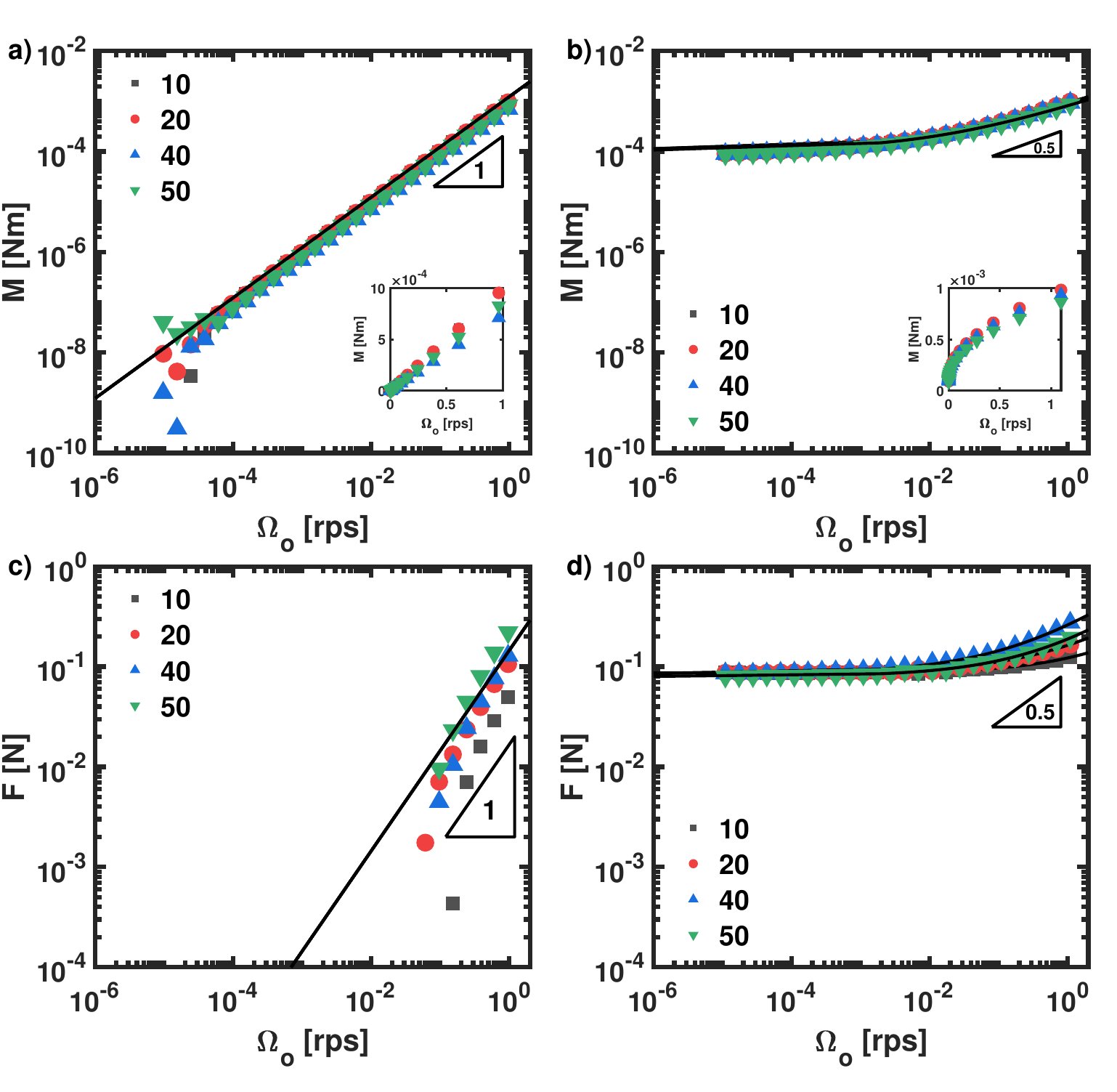}
\caption {Torque as a function of rotation rate for \textbf{a)} castor oil and \textbf{b)} Carbopol, measured using geometries with ridges at various $\theta$ (indicated in the legend). Torque scales linearly with rotation rate for castor oil (slope = 1), while for Carbopol, the torque follows a Herschel-Bulkley model relation with rotation rate  with $n \approx 0.5$. Insets in \textbf{a} and \textbf{b} show the variation of torque with rotation rate on a linear scale, where the geometric variations are more noticeable. Normal force as a function of rotation rate for \textbf{c)} castor oil  and \textbf{d)} Carbopol, measured using geometries with ridges at various $\theta$ (indicated in the legend). For castor oil, normal force exhibits linear scaling with rotation rate (slope~=~1), whereas in the case of Carbopol, normal force scales non-linearly with rotation rate in a Herschel-Bulkley fashion with $n \approx 0.5$. In all cases, $F>0$, indicating a force against gravity hence the downward displacement of the fluid by the geometry. The solid lines represent linear fits (in \textbf{a, c}) and the Herschel-Bulkley fits (in \textbf{b, d}), respectively.}
\label{fig:rheo_Newtonian_Carbopol}
\end{figure}

We present the flow curves of castor oil and Carbopol measured from the bulk rheological experiments. For castor oil, we  see a trivial linear dependence of both torque $M$ and normal force $F$ on the imposed rotation rate, as shown in Figures~\ref{fig:rheo_Newtonian_Carbopol}a and c, respectively (solid lines correspond to the linear fit). For Carbopol, both torque and normal force are well described by the Herschel-Bulkley model as a function of the imposed rotation rate, as depicted in Figures~\ref{fig:rheo_Newtonian_Carbopol}b and d, respectively. The solid lines indicate the Herschel-Bulkley model fit and the obtained flow indices are found to be around 0.5, which are approximately equal to the flow index obtained from the stress versus shear rate variation using a standard Couette geometry (as given in Appendix Figure~\ref{fig:shearrate_ramp_Carbopol_Emulsion_SI}). For both the fluids, all the flow curves (Figures~\ref{fig:rheo_Newtonian_Carbopol}a and b) appear to overlap when viewed on a typical log–log axes, however, the geometric variations are more observable on a linear scale, as illustrated in the insets of Figures~\ref{fig:rheo_Newtonian_Carbopol}a and b. 

The rotation of the angled geometry causes the velocity gradient to penetrate into the space between the ridges. This penetration generates a vertical pushing force on the fluid, resulting in a normal force, and consequently, an axial flow in the $z$-direction. The presence of the normal force and the axial flow indicates the existence of normal stress differences in shear flows, as previously reported in emulsions and foams,\cite{Habibi2016, deCagny2019} as well as in polymer solutions and suspensions.\cite{Gauthier2021} However, due to the complexity of the angled geometry, deriving expressions for normal stress differences in terms of normal force is challenging, unlike in simpler geometries such as cone-plate or parallel plate geometries.\cite{Bird1987} Therefore, to estimate the yield normal stress, we treated the geometry as a smooth solid cylinder without ridges, with $R_i=8.5$ mm, $R_o=9$ mm and $L=25.4$ mm. The estimated yield normal stress for Carbopol found in this way was $\sim$62~Pa.

The angle of the ridges plays a significant role in influencing the penetration depth of the fluid and the resulting axial flows. While the azimuthal flow appears largely unaffected by the ridge angle, the impact is more pronounced in the axial direction, as depicted in Figures~\ref{fig:rheo_Newtonian_Carbopol}c and d. For the two limiting cases, i.e., for vertical ridges ($\theta=0^\circ$) and horizontal ridges ($\theta=90^\circ$), the vertical pushing force is zero, leading to zero axial flow. Between these limits,  the larger the angle, the greater the vertical pushing force, which enhances the axial flow, reaching a maximum at an intermediate ridge angle. For smaller ridge angles, the geometry partially acts like a vane geometry, limiting the penetration of the fluid flow between the ridges and thus, decreasing the downward flow. At larger ridge angles, the fluid flow penetration is enhanced as shown in our previous work,\cite{Majhi2024} resulting in increased downward flow and a higher normal force. Furthermore, it is evident that changes in ridge angle result in only minor variations in yield normal stress, while the viscous effects in the axial direction exhibit a more pronounced dependence on the ridge angle (see Figure~\ref{fig:rheo_Newtonian_Carbopol}d). This directional dependence suggests that the angled ridged geometry primarily influences the normal force and axial flow, while its effect on the rotational (azimuthal) flow remains minimal. To further explore the effect of geometric variations, we measured the local flow behaviour of all test fluids using rheo-MRI. The detailed results are presented in the following subsection.

\subsection{Local velocity profiles from rheo-MRI experiments}

To gain further insight into the local flow behaviour of fluids in angled ridged geometries, we measured the time- and azimuthal angle-averaged velocity profiles using rheo-MRI. These profiles were obtained in both the rotational ($xy$-plane) and axial ($z$-axis) directions. We present the velocity profiles across the gap for the three test fluids in ridged geometries with angles of 10$^\circ$, 20$^\circ$, and 45$^\circ$. In this subsection, we examine each test fluid individually. For each test fluid, we first discuss our observations for the rotational velocity profiles in different angled ridged geometries under various imposed rotation rates. We then highlight the differences in the corresponding axial velocity profiles under the same parametric variations. 

\subsubsection{Castor Oil}

\begin{figure}[h!]
\centering
\includegraphics[scale=0.28]{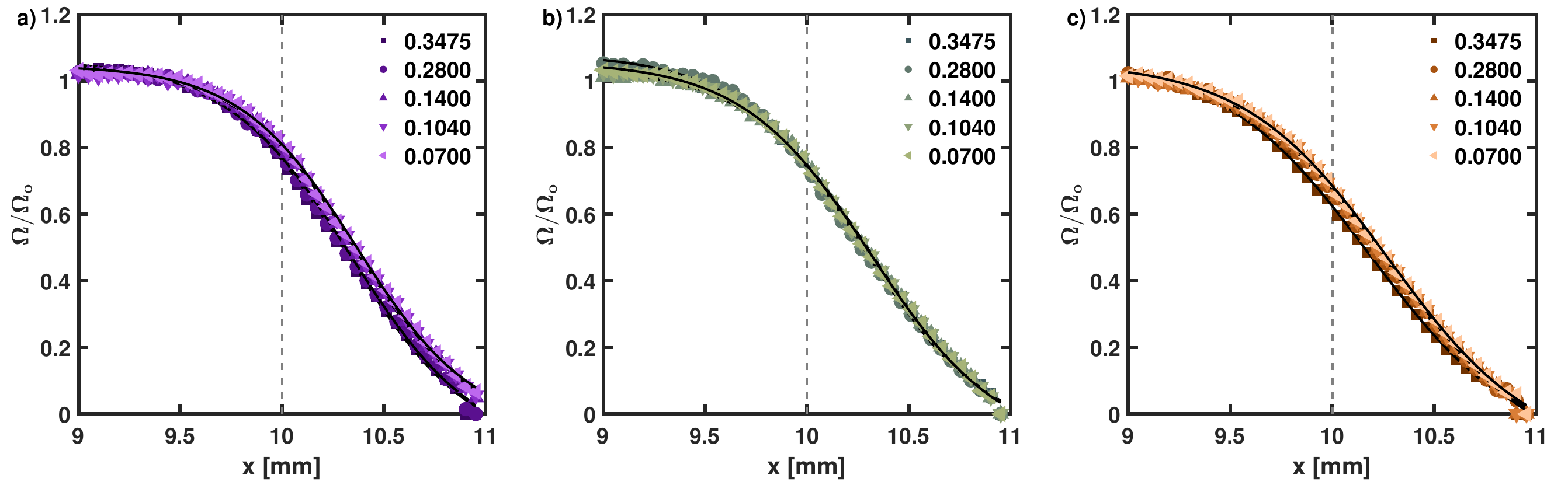}
\includegraphics[scale=0.28]{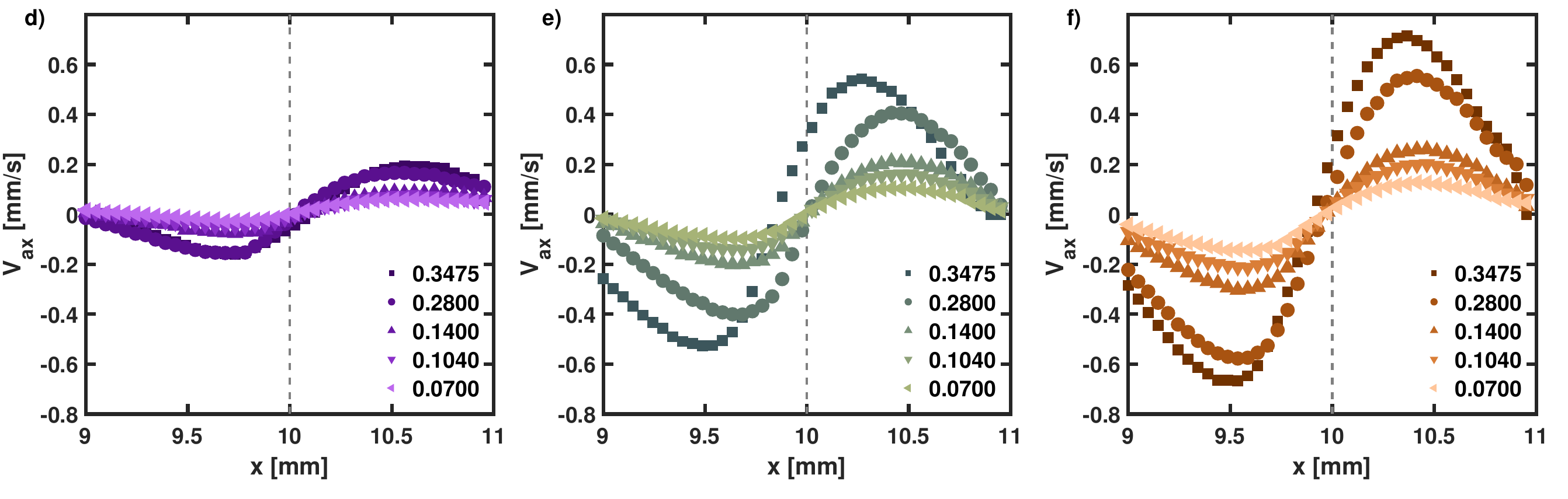}
\caption {Rotational velocity \textbf{(a--c)}, normalised with respect to the imposed rotation rates and axial velocity \textbf{(d--f)} as a function of position in the gap for castor oil, measured at different imposed rotation rates. Each panel presents velocities in geometries with ridges at three different angles: purple for 10$^\circ$, green for 20$^\circ$, and brown for 45$^\circ$. Legends denote the specific imposed rotation rates, $\Omega_\textnormal{o}$ in rps, for each curve, while the dashed line marks the tip of the ridge. The solid lines are fits to a sigmoidal function (Equation~(\ref{ch3:eq:sigmoidalfunction})), from which the penetration depths are obtained.}
\label{fig:velprof_Newtonian}
\end{figure}

The rotational velocity profiles for castor oil at different imposed rotation rates $\Omega_\textnormal{o}$ are presented in panels a--c of Figure~\ref{fig:velprof_Newtonian_SI}. Each panel corresponds to ridged geometries with angles of 10$^\circ$, 20$^\circ$, and 45$^\circ$, respectively. These profiles clearly demonstrate that the fluid at the inner wall moves with an identical rotational velocity as the inner geometry and reference fluid, indicating that there is no wall slip. Furthermore, as the imposed rotation rate increases, the magnitude of the rotational velocity across the gap also increases. 
To compare the rotational velocity profiles across different imposed rotation rates, we normalise the rotational velocity, $\Omega$, by the respective imposed rotation rate $\Omega_\textnormal{o}$. As expected due to the linearity of the Stokes equation, this normalisation collapses the data on a master curve (Figures~\ref{fig:velprof_Newtonian}a--c).
The velocity profiles exhibit a plateau near the inner wall up to a specific $x$ value, followed by a gradual decline towards zero as $x$ approaches the outer wall, in accordance with the no-slip boundary condition at the outer wall. This plateau occurs because, near the inner wall, the fluid in the gap between the ridges exhibits solid body rotation, rotating in sync with the inner cylinder. Clearly, however, the fluid velocity declines already before the tip of the ridges, confirming that the velocity gradient penetrates into the region between the ridges. 

Interestingly, the extent of the initial plateau region varies with ridge angle, increasing with smaller ridge angles, with the plateau extending to the largest $x$ value for the smallest angle. The increased plateau region with smaller ridge angles is attributed to the fact that ridges at smaller angles cause less disruption in the fluid flow, allowing the fluid to maintain solid body rotation over a larger portion of the gap, thereby, extending the plateau region further into the gap. Conversely, with ridges at larger angles introduce greater resistance, limiting the extent of the plateau region. This is in line with our previous findings (see ref.~\citenum{Majhi2024}), where we found that the penetration of the velocity gradient in the region between the ridges is smaller for smaller angles. Later in Subsection~\ref{comp_analysis_velprof}, we will analyse the penetration of the fluid flow more quantitatively.

Having analysed the rotational velocity profiles, we now turn our attention to the axial velocity profiles, which provide crucial insights into the secondary flow behaviour induced by the angled ridges. The axial velocity profiles for castor oil at different imposed rotation rates $\Omega_\textnormal{o}$ are presented in panels d--f of Figure~\ref{fig:velprof_Newtonian}. Each panel corresponds to ridged geometries with angles of 10$^\circ$, 20$^\circ$, and 45$^\circ$, respectively. As previously discussed, the penetration of the fluid flow between the ridges creates a vertical pushing force on the fluid that induces an axial flow in the $z$-direction. In the gap within the ridges, the fluid flows downward while an equivalent mass flux of fluid moves upward in the gap between the ridge tip and the outer cylinder. We observe that the magnitude of the axial velocity increases with the imposed rotation rate across the entire gap, regardless of the ridge angle. This rate dependence of the axial velocities is evidently a direct consequence of the linearity of the Stokes equation. Larger ridge angles strongly enhance the axial velocity, consistent with the higher normal forces measured in geometries with larger ridge angles. This enhancement results from the greater disruption to the flow caused by large ridge angles, which pushes more fluid vertically into the gap, ultimately increasing the axial velocity.

Having established the reference velocity profiles for a Newtonian fluid using castor oil, we then proceed to discuss the more complex velocity profiles of yield stress fluids, starting with Carbopol and subsequently moving on to the emulsion. 

\subsubsection{Carbopol}

\begin{figure}[h!]
\centering
\includegraphics[scale=0.28]{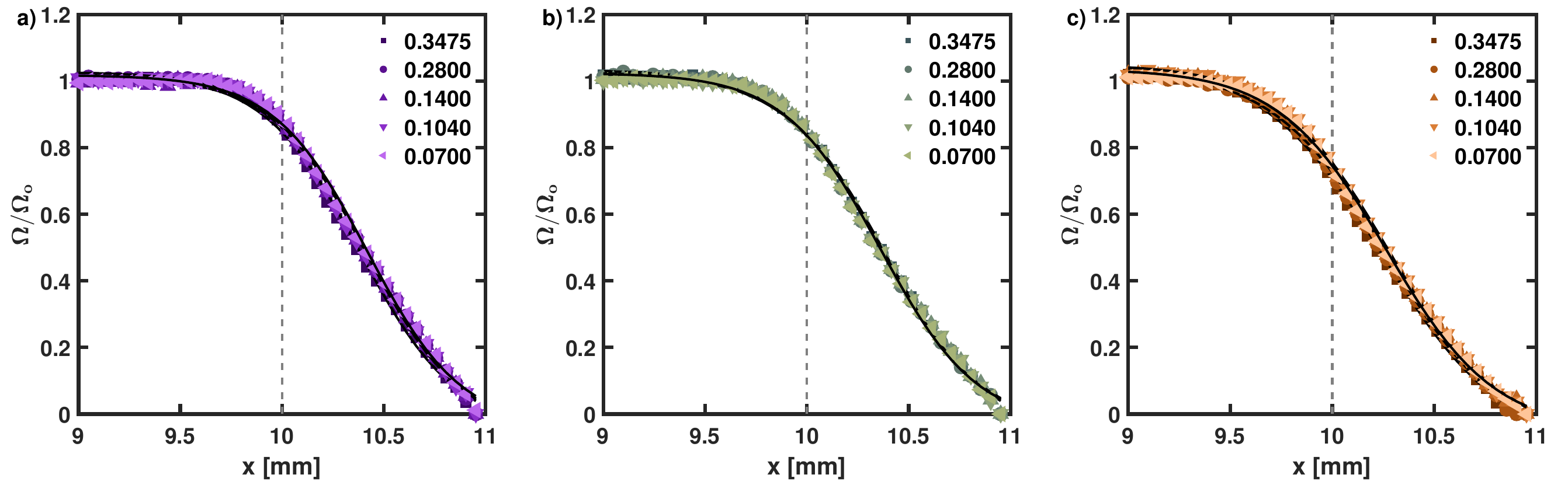}
\includegraphics[scale=0.28]{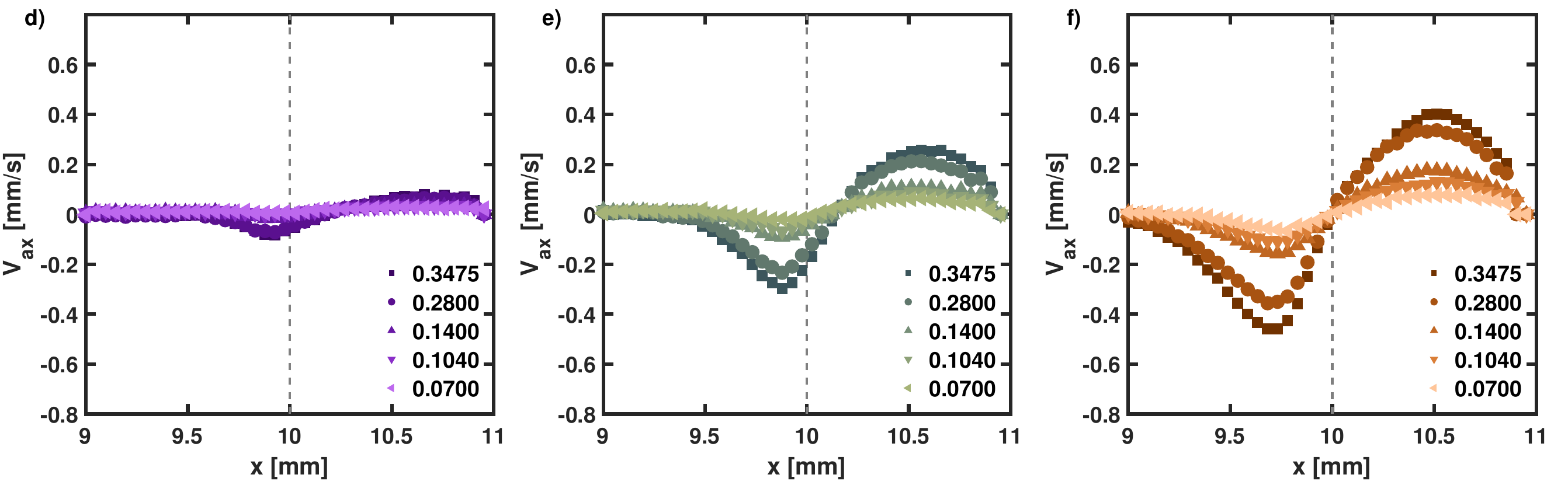}
\caption {Rotational velocity \textbf{(a--c)}, normalised with respect to the imposed rotation rates and axial velocity \textbf{(d--f)} as a function of position in the gap for Carbopol, measured at different imposed rotation rates. Each panel presents velocities in geometries with ridges at three different angles: purple for 10$^\circ$, green for 20$^\circ$, and brown for 45$^\circ$. Legends denote the specific imposed rotation rates, $\Omega_\textnormal{o}$ in rps, for each curve, while the dashed line marks the tip of the ridge. The solid lines are fits to a sigmoidal function (Equation~(\ref{ch3:eq:sigmoidalfunction})), from which the penetration depths are obtained.}
\label{fig:velprof_Carbopol}
\end{figure}

The normalised rotational velocity profiles for Carbopol at different imposed rotation rates $\Omega_\textnormal{o}$ are shown in panels a--c of Figure~\ref{fig:velprof_Carbopol}, while the corresponding non-normalised profiles are presented in panels a--c of Figure~\ref{fig:velprof_Carbopol_SI}. Each panel corresponds to ridged geometries with angles of 10$^\circ$, 20$^\circ$, and 45$^\circ$, respectively. Similar to castor oil, these profiles exhibit no wall slip at both the inner and outer walls, and the magnitude of the rotational velocity across the gap increases with the imposed rotation rate, irrespective of the ridge angle. The profiles also display a plateau region near the inner wall, indicative of solid body rotation, followed by a decline towards the outer wall. However, the plateau extends further into the gap for Carbopol than for castor oil, because of the reduced penetration of the fluid flow into the gap within the ridges (we will analyse this more quantitatively later in Subsection~\ref{comp_analysis_velprof}). This indicates that the yield stress acts as a barrier for further penetration, causing a larger portion of the fluid to move as a rigid body. The extent of the initial plateau region again varies with ridge angle: the plateau is more evident at smaller ridge angles, extending to greater $x$ values as the ridge angle decreases. 

The normalised velocity profiles in Figures~\ref{fig:velprof_Carbopol}a--c collapse on a master curve. This is somewhat surprising, as we showed that the Carbopol suspensions have a non-linear relation between the shear stress and shear rate. In particular, one might expect that the presence of a yield stress would lead to a stress (and therefore rotation rate)-dependent penetration depth, as the penetration of the fluid flow between the ridges requires the material to yield. Apparently, this effect is very small for Carbopol in the range of rotation rates studied, which might be due to the relatively small value of the yield stress in this system. Most likely, the shear stresses induced by the applied rotation far exceed the yield stress, making any effects of the yield stress on the flow profile relatively small. We note that a similar collapse of velocity profiles was observed in a study of sheared dense suspensions at rotation rates that are in the shear rate-dependent part of the flow curves.\cite{Ovarlez2006}

The axial velocity profiles for Carbopol at different imposed rotation rates $\Omega_\textnormal{o}$ are presented in panels d--f of Figure~\ref{fig:velprof_Carbopol}. Each panel corresponds to ridged geometries with angles of 10$^\circ$, 20$^\circ$, and 45$^\circ$, respectively. These profiles show several notable differences when compared to castor oil. The magnitude of the axial velocities for Carbopol is lower than that for castor oil (this trend will be more apparent in the next subsection, where the maximum axial velocities observed across all the test fluids are compared). This reduction in the axial velocity is due to the reduced penetration of the fluid flow into the gap within the ridges in Carbopol relative to castor oil. Additionally, the axial velocities for Carbopol are found to increase with both the imposed rotation rate (due to increased flow) and the ridge angle (due to greater flow disruption). 

Beyond the difference in magnitude, the shape of the axial velocity profiles for Carbopol also differs significantly from the nearly symmetric profiles observed in castor oil. Specifically, the zero-crossing point from negative ${V_\textnormal{ax}}$ to positive ${V_\textnormal{ax}}$ is slightly shifted away from the ridge tip and towards the outer wall. This shift may be related to the shear banding typically observed in the rotational velocities of yield stress fluids near the ridge tip.  Additionally, unlike the gradual variation in axial velocity profiles seen across the gap in castor oil, the profiles for Carbopol show a region near the inner cylinder where the axial velocity is close to zero, indicating reduced flow profile penetration. This zero axial velocity region is influenced by both the imposed rotation rate  and the ridge angle, decreasing with increasing imposed rotation rate as well as with larger ridge angles. The reduction in the extent of the zero axial velocity region is due to the resulting higher local shear rates in the axial direction, which arise from the increased axial flow at higher imposed rotation rates and in geometries with large ridge angles. 

Building upon the insights gained from the velocity profiles for Carbopol, we now extend our analysis to the castor oil-in-water emulsion, another yield stress fluid, but with a higher yield stress.

\subsubsection{Castor oil-in-water emulsion}

\begin{figure}[h!]
\centering
\includegraphics[scale=0.28]{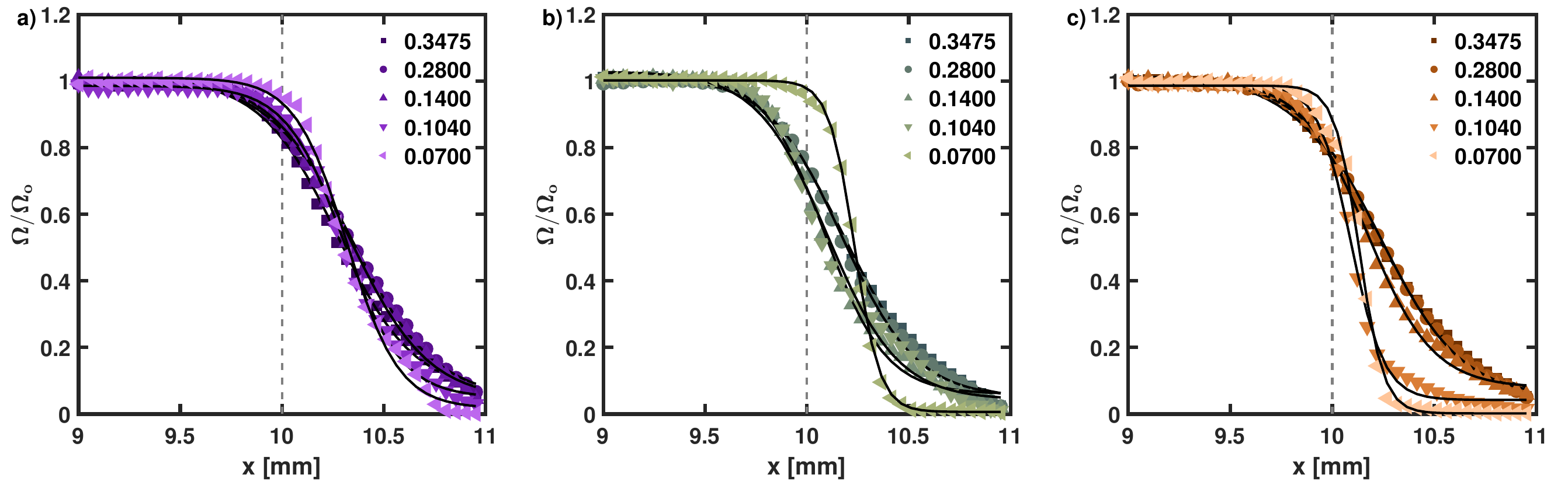}
\includegraphics[scale=0.28]{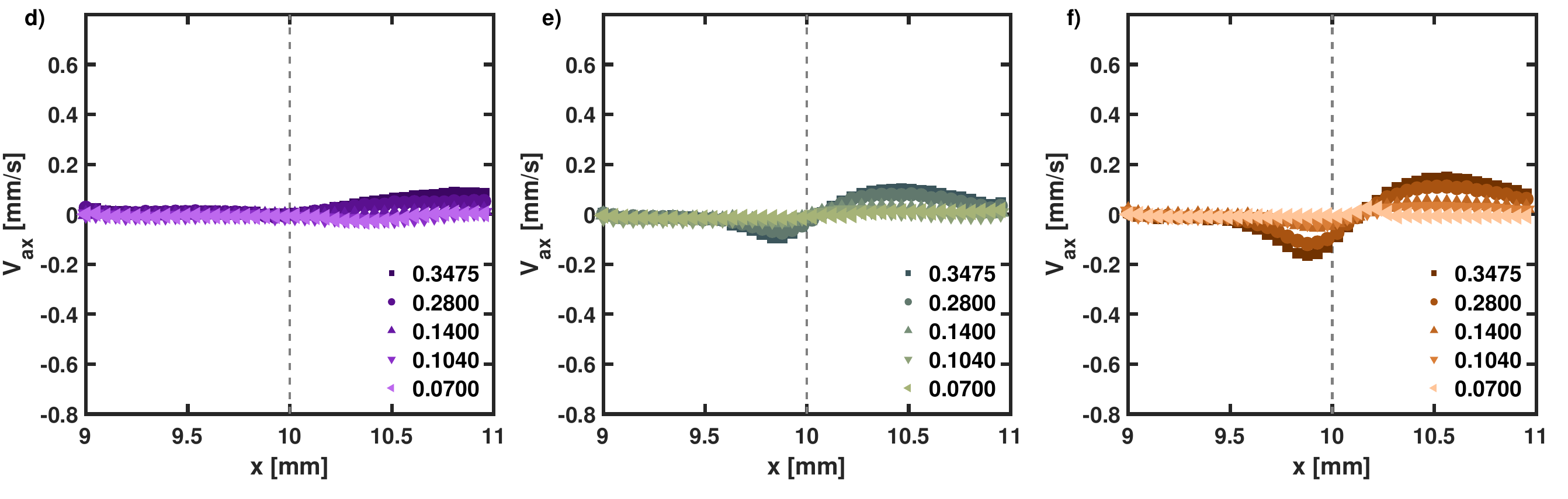}
\caption {Rotational velocity \textbf{(a--c)}, normalised with respect to the imposed rotation rates and axial velocity \textbf{(d--f)} as a function of position in the gap for castor oil-in-water emulsion, measured at different imposed rotation rates. Each panel presents velocities in geometries with ridges at three different angles: purple for 10$^\circ$, green for 20$^\circ$, and brown for 45$^\circ$. Legends denote the specific imposed rotation rates, $\Omega_\textnormal{o}$ in rps, for each curve, while the dashed line marks the tip of the ridge. The solid lines are fits to a sigmoidal function (Equation~(\ref{ch3:eq:sigmoidalfunction})), from which the penetration depths are obtained.}
\label{fig:velprof_Emulsion}
\end{figure}

The normalised rotational velocity profiles for castor oil-in-water emulsion at different imposed rotation rates $\Omega_\textnormal{o}$ are shown in panels a--c of Figure~\ref{fig:velprof_Emulsion}, while the corresponding non-normalised rotational velocity profiles are presented in panels a--c of Figure~\ref{fig:velprof_Emulsion_SI}. Each panel corresponds to ridged geometries with angles of 10$^\circ$, 20$^\circ$, and 45$^\circ$, respectively. 
Clearly, there is a key difference in these profiles compared to castor oil or Carbopol: the normalised rotational velocity profiles do not collapse onto a single master curve. This failure to collapse is due to the non-linear scaling of the rotational velocities with the imposed rotation rate. 

Beyond the global difference in the behaviour, the rotational velocity profiles for the emulsion share some qualitative similarities with those for Carbopol. They demonstrate no wall slip at both the inner and outer walls, and feature a plateau region indicative of solid body rotation near the inner wall, followed by a decrease in the velocity approaching zero at the outer wall. 

In comparison to Carbopol, the effect of the yield stress is more pronounced in the emulsion; consequently, the flow profile penetration into the gap between the ridges is further suppressed. Notably, for the two smallest ridge angles, flow penetration ceases entirely below a certain imposed rotation rate, as evident from the flat plateau region extending to the ridge tip. As a result, the emulsion exhibits the most prominent plateau region among all the tested fluids, a direct consequence of its higher yield stress. The plateau region in the emulsion, as in Carbopol, is more extended at smaller ridge angles. Conversely, this plateau region is significantly influenced by the imposed rotation rate in the emulsion, and this difference is most noticeable at lower imposed rotation rates.

In the gap outside the ridges, the normalised rotational velocity profiles appear more curved than for castor oil and Carbopol, as also previously reported in studies on silicone oil-in-water emulsion,\cite{Hollingsworth2004, Ovarlez2008} mayonnaise,\cite{Hollingsworth2004} micronised fat crystals in oil\cite{Nikolaeva2018} and egg yolk emulsions.\cite{Serial2022} This curvature arises because the gap in our experimental setup, with an inner to outer radii ratio of 0.82, induces a considerable stress variation across the gap. Quantitatively, there is a 33$\%$ drop in the local shear stress across the gap. While this stress decay is small enough to maintain an almost linear velocity profile for a Newtonian fluid, it significantly impacts the velocity profiles for a yield stress fluid, resulting in a more curved profile.\cite{deKort2018} Additionally, at the two lowest imposed rotation rates, shear localisation is observed\cite{Coussot2009, Ovarlez2008, Ovarlez2009, Becu2006, Paredes2011, Nikolaeva2018} in the emulsion, where the velocity tends to zero at some distance before the outer cylinder, and the fluid near the outer cylinder remains stagnant, suggesting that the local shear stress drops below the yield stress at this radial position.\cite{Hollingsworth2004} This shear localisation (or shear banding) effect is more prominent at larger ridge angles. At the higher imposed rotation rates, flow extends throughout the gap outside the ridges; the local shear stress at the inner cylinder is high enough that, even at the outer cylinder, where it is 67$\%$ of the local shear stress at the inner cylinder, the local stress still exceeds the yield stress.

Given the restricted penetration of the flow profile between the ridges for the emulsion and its effect on the rotational velocity profiles, we now proceed to discuss the axial velocity profiles for castor oil-in-water emulsion at various imposed rotation rates $\Omega_\textnormal{o}$, as displayed in panels d--f of Figure~\ref{fig:velprof_Emulsion}. Each panel corresponds to ridged geometries with angles of 10$^\circ$, 20$^\circ$, and 45$^\circ$, respectively. Compared to castor oil and Carbopol, the magnitude of the axial velocity is lowest for the emulsion (as we will discuss more quantitatively later in Subsection~\ref{comp_analysis_velprof}), a trend consistent with the reduced penetration of the fluid flow into the gap within the ridges. Similar to Carbopol, the axial velocity for the emulsion is observed to increase with both the imposed rotation rate and the ridge angle, reflecting the influence of increased flow and greater flow disruption, respectively. Like Carbopol, the axial velocity profiles for the emulsion are asymmetric around the ridge tip, and we notice that the zero-crossing point, where the axial velocity transitions from negative to positive, does not occur exactly at the ridge tip but rather at a location between the ridge tip and the outer wall. The shift in the zero-crossing point is likely due to the observed shear banding in the rotational velocities for the emulsion around the ridge tip (see top panel in Figure~\ref{fig:velprof_Emulsion}). Furthermore, the axial velocity profiles drop to zero at some position within the gap between the inner cylinder and the tip of the ridge. Consequently, close to the inner cylinder, we do not see any axial velocity, in particular for small ridge angles and low imposed rotation rates.

\subsection{Quantitative analysis of the velocity profiles}
\label{comp_analysis_velprof}

In the previous subsection, we presented the velocity profiles in both the azimuthal and axial directions for the different fluids in various angled ridged geometries. In this subsection, we analyse these velocity profiles in more detail to make a quantitative comparison between the three different fluids. In our previous work, reported in ref.~\citenum{Majhi2024}, we introduced the penetration depth $\delta$ as a measure for how much the velocity gradient penetrates in the region between the ridges. This penetration depth was obtained for a Newtonian fluid from the measured torque versus rotation rate, from which an effective gap size could be determined (i.e. the gap size in a geometry with an equivalent smooth inner cylinder leading to the same torque). In the present work, we obtain the penetration depth directly    
from the measured velocity profiles. Additionally, from the axial velocity profiles, we determine the maximum axial velocity within the ridges. The method for calculating the penetration depth and the maximum axial velocity is described in Subsection~\ref{subsec:calcmethod_delta_maxVax}. We explore the correlation between penetration depth, maximum axial velocity, and fluid type, specifically addressing how these parameters vary with different ridge angles and imposed rotation rates for the three test fluids.

\subsubsection{Penetration depth analysis}

\begin{figure}[b!]
\includegraphics[scale=0.25]
{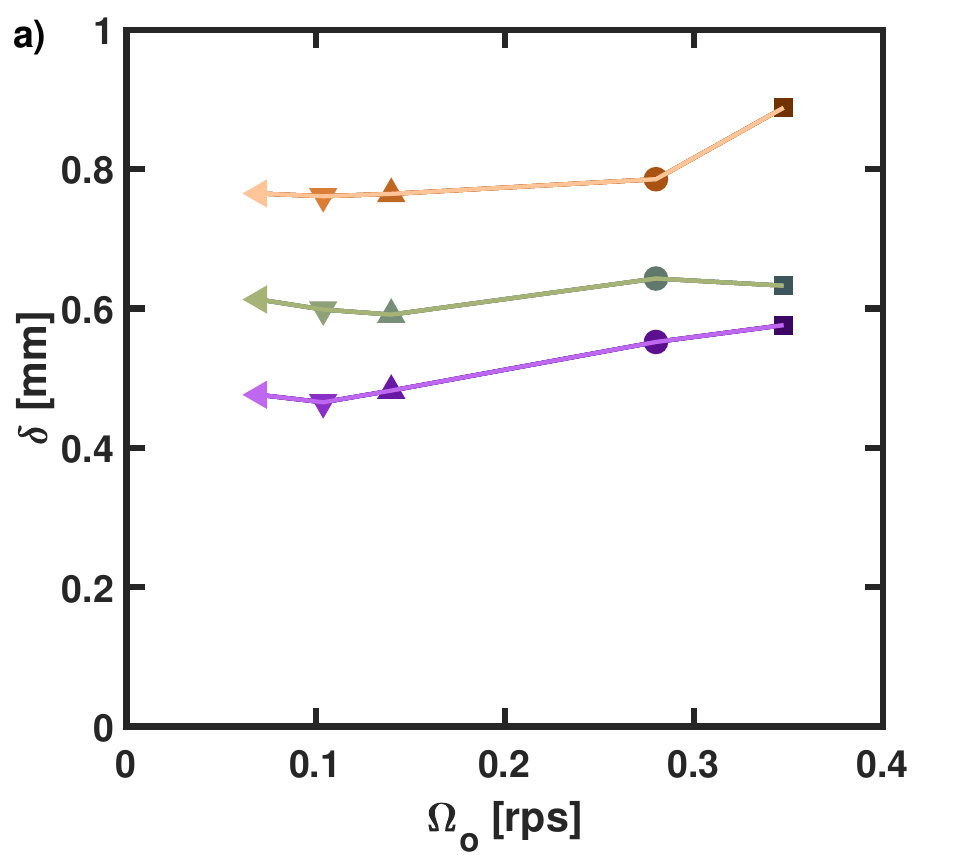}
\includegraphics[scale=0.25]
{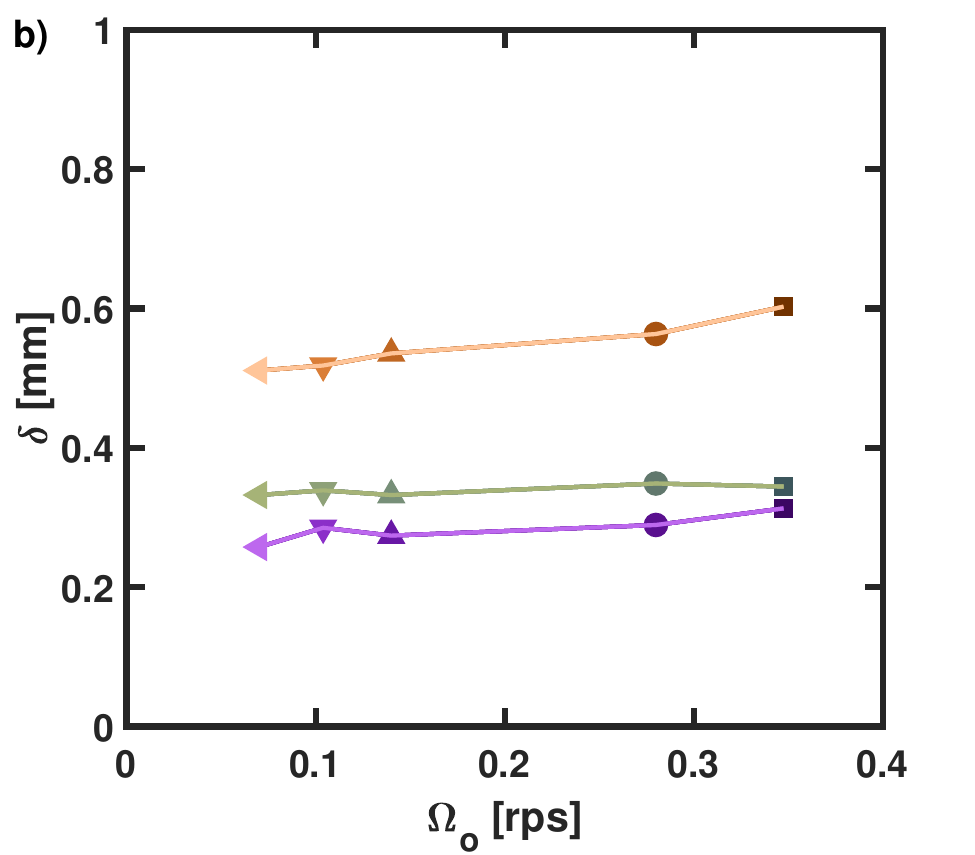}
\includegraphics[scale=0.25]
{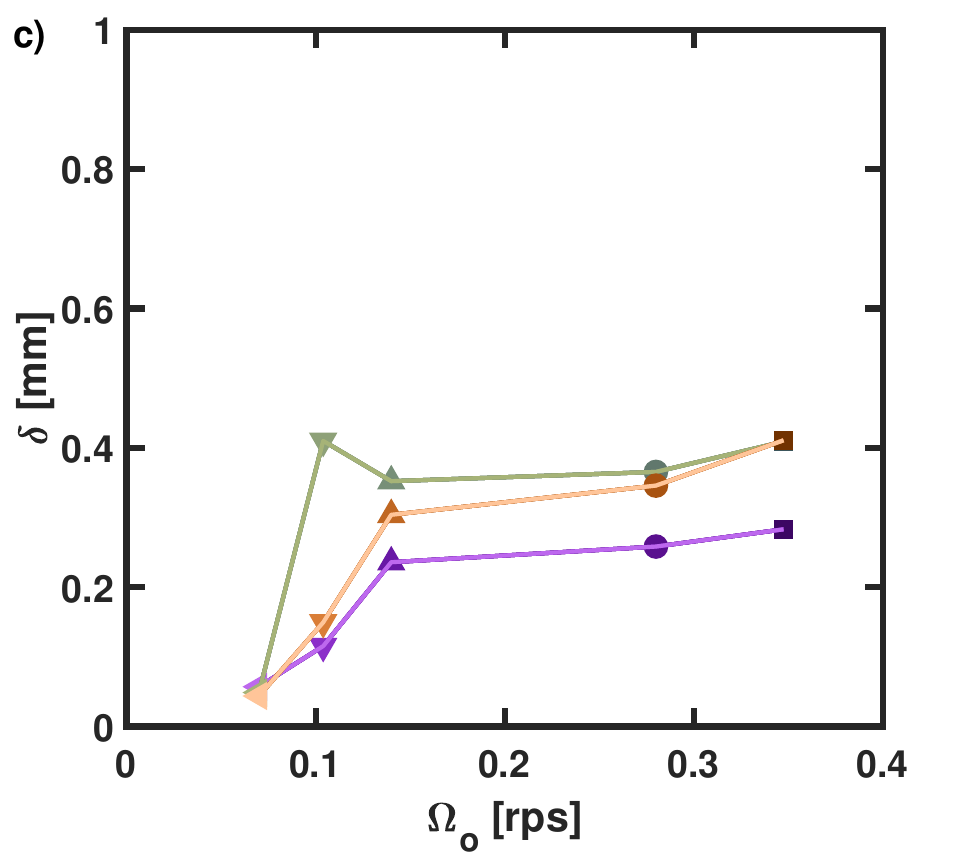}
\caption {Penetration depth $\delta$ as a function of imposed rotation rate in geometries with ridges at three different angles for \textbf{a)}~castor oil, \textbf{b)} Carbopol and \textbf{c)} castor oil-in-water emulsion. The coloured symbols represent different ridge angles: purple for 10$^\circ$, green for 20$^\circ$, and brown for 45$^\circ$ while the lines are guides to the eye.}
\label{fig:delta_all_fluids}
\end{figure}

The variation of the penetration depth with the imposed rotation rate is plotted in Figure~\ref{fig:delta_all_fluids}. Each panel shows data for geometries with varying ridge angles and represents the three test fluids. For all ridge angles and rotation rates, the penetration depth is the largest for castor oil, followed by Carbopol, and is smallest for the emulsion. This trend indicates that the fluid flow penetration in the gap within the ridges is higher for a Newtonian fluid compared to  yield stress fluids. Furthermore, the larger the yield stress of the fluid, the lesser is the flow penetration, leading to smaller values of $\delta$. Within each fluid, $\delta$ increases with ridge angle; geometries with ridges at 45$^\circ$ exhibit the largest $\delta$, followed by 20$^\circ$, and then 10$^\circ$ angles. This result is consistent with our findings from previous work.\cite{Majhi2024} However, an exception is observed for the emulsion, where the 20$^\circ$ geometry shows a larger $\delta$ than the 45$^\circ$ geometry. This anomaly may be attributed to the interpretation method used to identify the drop-off position or the potential limitations in the choice or fitting approach of the sigmoidal function.

Regarding the dependence on rotation rate, for Newtonian fluids such as castor oil, $\delta$ should theoretically remain constant, as the shape of the flow pattern is independent of the imposed rotation rate in the Stokes limit. However, the experimental $\delta$ values for castor oil show a slight dependence on the rotation rate, which could be due to subtle effects not captured by the idealised theoretical model or uncertainties in determining the drop-off position.

In case of Carbopol, the penetration depth varies weakly with the imposed rotation rate, similar to castor oil. By contrast, for the emulsion, the penetration depth increases sharply at low imposed rotation rates, followed by a more gradual increase at higher imposed rotation rates. The smaller $\delta$ values at low rotation rates are caused by the yield stress, which suppresses penetration of the flow into the ridged geometry. At sufficiently low imposed rotation rates, the complete region between the ridges remains unyielded, reducing $\delta$ to zero. As the imposed rotation rate increases, the emulsion exhibits more extensive yielding behaviour, causing $\delta$ to increase.

\subsubsection{Axial flow analysis}

\begin{figure}[t!]
\includegraphics[scale=0.25]
{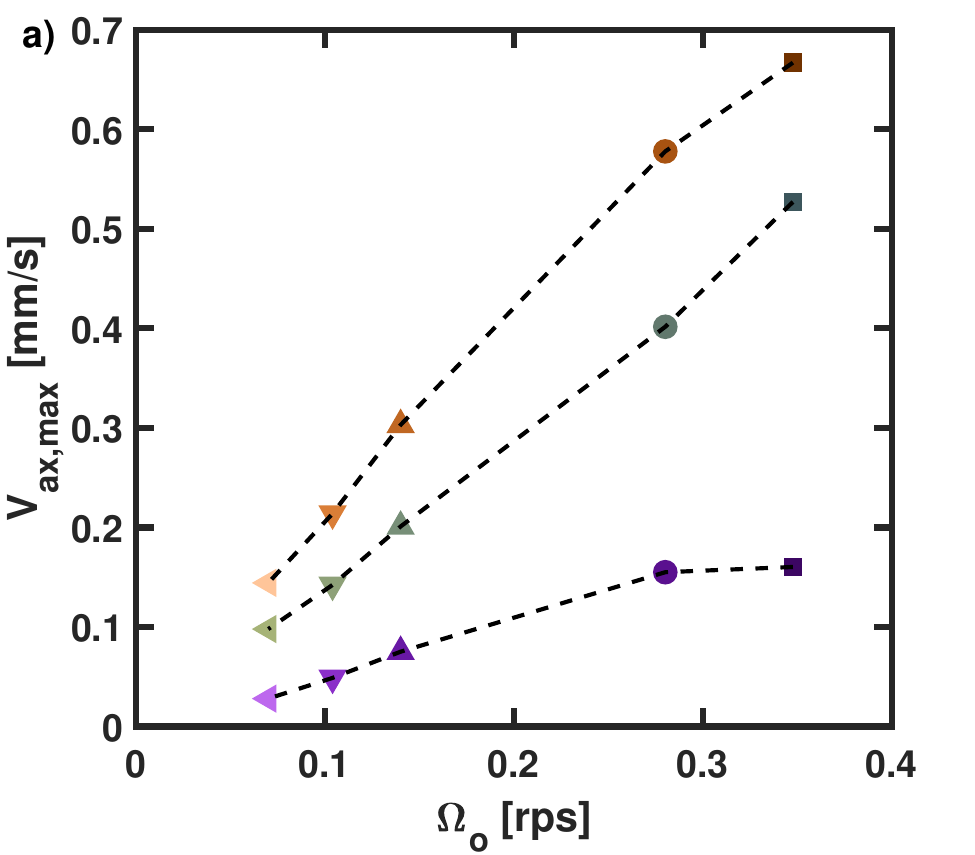}
\includegraphics[scale=0.25]
{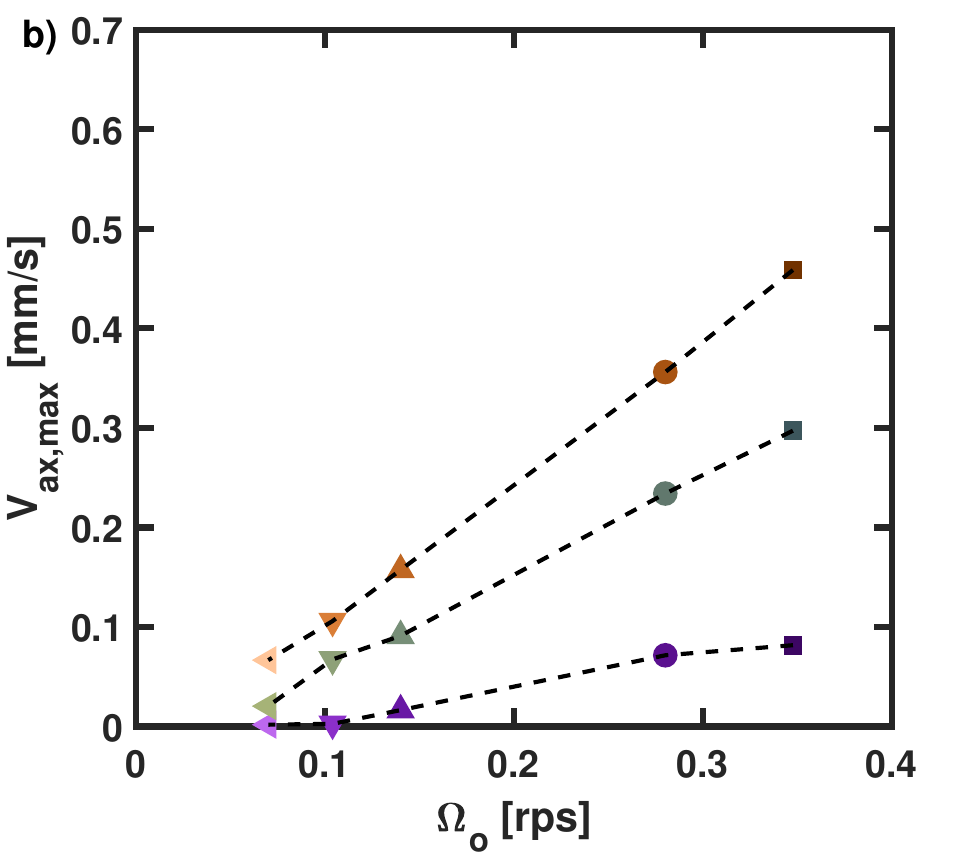}
\includegraphics[scale=0.25]
{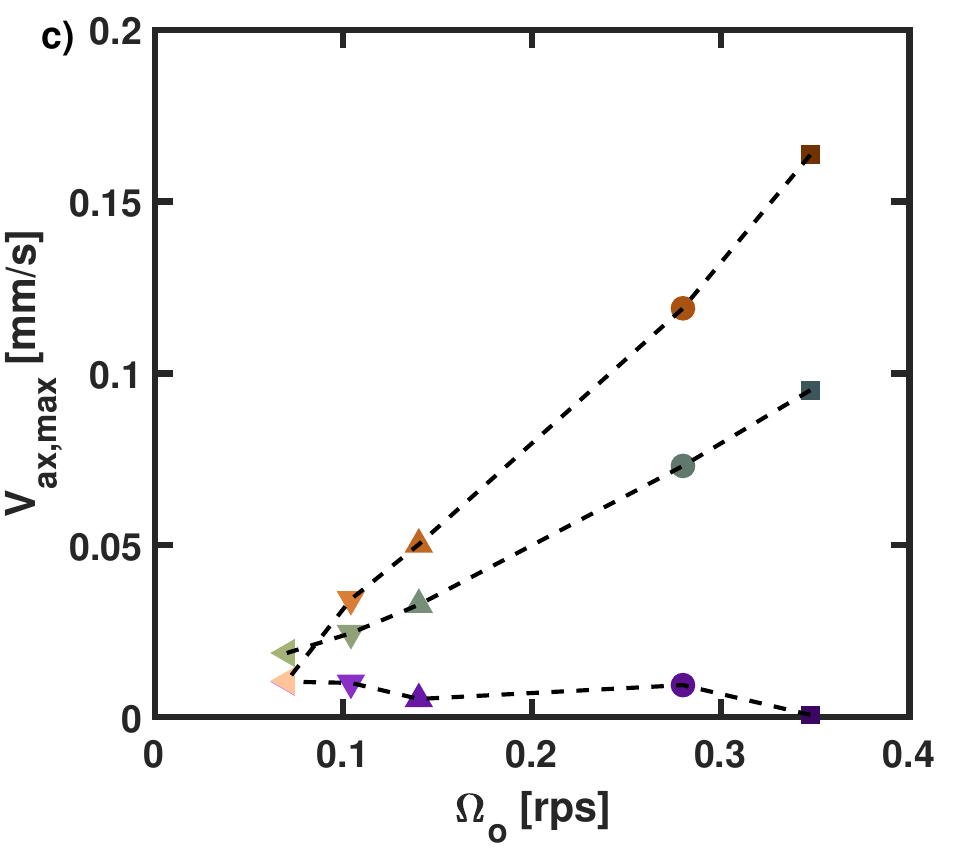}
\caption {Maximum axial velocity as a function of imposed rotation rate in geometries with ridges at three different angles for \textbf{a)}~castor oil, \textbf{b)} Carbopol and \textbf{c)} castor oil-in-water emulsion. The coloured symbols indicate different ridge angles: purple for 10$^\circ$, green for 20$^\circ$, and brown for 45$^\circ$. The black dashed lines are guides to the eye to illustrate the linear trend of the data. In~\textbf{a)},~maximum $V_\textnormal{ax}$ = 0 at an imposed rotation rate of zero, indicating no axial flow when there is no rotation. In~\textbf{b)}~and \textbf{c)}, the maximum axial velocity scales linearly with an offset, indicating that while there is flow in the primary direction, axial flow is absent, as evidenced by the intercept on the rotation rate axis. Note the different scale on the vertical axis in \textbf{c} compared to \textbf{a} and \textbf{b}.}
\label{fig:maxVax_all_fluids}
\end{figure}

The variation of the maximum axial velocity with the imposed rotation rate is plotted in Figure~\ref{fig:maxVax_all_fluids}. Again, the data is shown for geometries with varying ridge angles and the three test fluids. The variation in the penetration depth with the imposed rotation rate directly impacts the corresponding axial flow. For all ridge angles and rotation rates, the maximum axial velocities are highest for castor oil, followed by Carbopol, and are lowest for the emulsion. Additionally, within each fluid, the maximum axial velocity increases as ridge angle increases. This variation correlates with the increased flow field penetration into the ridged geometry in case of castor oil and in geometries with larger ridge angles, in comparison to the two yield stress fluids and geometries with ridges at smaller angles, respectively.  

While the maximum axial velocities for castor oil are proportional to the applied rotation rate, as expected for a Newtonian fluid (Figure~\ref{fig:maxVax_all_fluids}a), for Carbopol and the emulsion, $V_\textnormal{ax,max}$ shows linear scaling with an offset (Figures~\ref{fig:maxVax_all_fluids}b and c): there appears to be a critical rotation rate below which there is no axial flow for the yield stress fluids. This is most likely related to the strong decline in the penetration depth at low rotation rates seen in Figure~\ref{fig:delta_all_fluids}c. 

\begin{figure}[h!]
\centering
\includegraphics[scale=0.5]{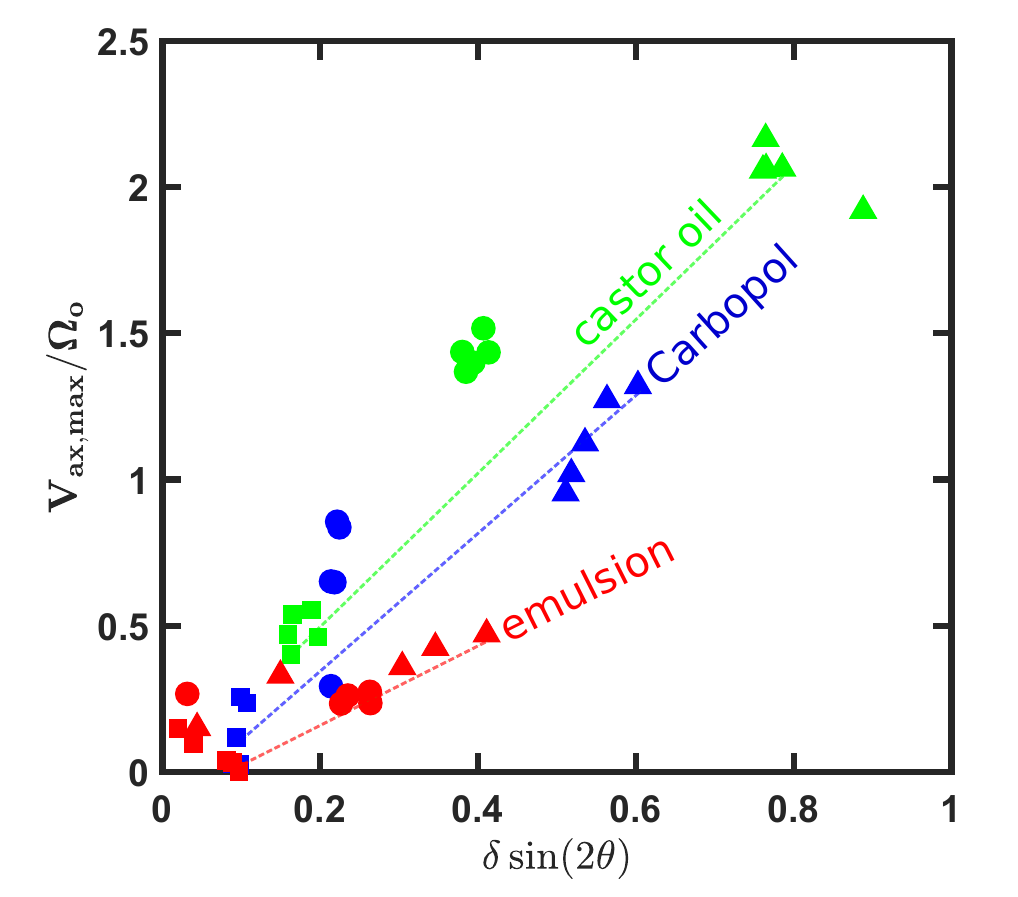}
\caption {The rescaled maximum axial velocity $V_\textnormal{ax,max}/\Omega_\textnormal{o}$ as a function of the penetration depth $\delta$ multiplied with $\sin(2\theta)$ for all three fluids. The different colours indicate different fluids, and the symbols are for the three different ridge angles: 10$^\circ$ (squares), 20$^\circ$ (circles), and 45$^\circ$ (triangles), respectively. The dashed lines are guides to the eye.}
\label{fig:maxVax_delta_all_fluids}
\end{figure}

As we argue that the axial flow is driven by penetration of the fluid flow between the ridges, we expect the maximum axial velocity to be proportional to the penetration depth~$\delta$. Furthermore, the vertical push given to the fluid by the ridges is expected to depend on the imposed rotation rate~$\Omega_\textnormal{o}$ and on the angle of the ridges~$\theta$. For vertical ridges ($\theta=0^\circ$) and horizontal ridges ($\theta=90^\circ$), there is no vertical component of the force exerted by the ridges, so we expect $V_\textnormal{ax,max}=0$ for these cases. A reasonable assumption is then that the axial velocity will vary as $V_\textnormal{ax,max}\sim\Omega_\textnormal{o}\delta\sin(2\theta)$. To verify this, we plot $V_\textnormal{ax,max}/\Omega_\textnormal{o}$ as a function of $\delta\sin(2\theta)$ in Figure~\ref{fig:maxVax_delta_all_fluids} for all three fluids. We see that for all three fluids, indeed a more or less linear relation is found. However, the slope of the lines is different. It is largest for the castor oil and smallest for the emulsion, which suggests that the yield stresses suppresses the axial flow, even for the same penetration depth of the shear flow. This highlights that the complete flow picture is not only determined by the flow profile penetration but also by the rheological properties of the fluid. In particular, it seems that for Carbopol, there is a combination of ridge angle and imposed rotation rate, for which there is a zero axial velocity, while there is still a finite penetration depth. This behaviour indicates the presence of a critical rotation rate, at which fluid in the gap between the ridges, is yielded in rotational direction while it is not yielded in the axial direction, contrary to the work of Ovarlez~\textit{et~al.},\cite{Ovarlez2010} where they showed that the imposed flow in one direction should always unjam the fluid flow in the secondary direction.

\section{Conclusions}

In this work, we investigated the flow behaviour of fluids in multi-directional flows using custom-designed patterned Couette geometries with ridges at an angle relative to the vertical axis. The bulk flow behaviour of a Newtonian fluid and two yield stress fluids was first examined in these angled geometries using conventional rheological measurements. Additionally, to capture the local flow behaviour, we measured the velocity profiles in both rotational and axial directions using rheo-MRI.

The rotation of the angled geometry causes the velocity gradient to penetrate into the space between the ridges. This penetration generates a vertical pushing force on the fluid, resulting in an axial fluid flow and a measurable normal force. Furthermore, the velocity profiles enabled direct measurement of the penetration depth and the maximum axial velocity. These two parameters increased systematically with ridge angles and imposed rotation rates across all three test fluids. Notably, a higher penetration depth correlated with increased axial flow, indicating that the flows in the azimuthal and axial directions are interconnected. Additionally, we identified a critical imposed rotation rate for yield stress fluids, below which the axial flow was completely suppressed.  

This study establishes a strong link between global rheological measurements and local flow behaviour in such complex geometries. It reveals the intricate interplay between the topology of the walls, imposed rotation rate, and the rheological properties of the fluid, particularly highlighting how the penetration depth and the maximum axial velocity are influenced by each other. These findings not only advance our fundamental knowledge of fluid behaviour in complex geometries but also have potential implications for various industrial applications where complex geometries and diverse fluid types are encountered.

\begin{acknowledgments}

The authors thank Yuri Hendrix (University of Amsterdam) for support with emulsion preparation; and Camilla Terenzi (Laboratory of Biophysics, Wageningen University and Research) for initial discussions and useful feedback.

The authors thank all the industrial partners and collaborators for their useful comments. This work is a part of the Industrial Partnership Programme Controlling Multiphase Flow (Project No. WP-30-01) that is carried out under an agreement between Shell, Unilever Research and Development B.V., Evodos and the Netherlands Organisation for Scientific Research (NWO). This project is co-funded by NWO and TKI-E$\&$I with the supplementary grant `TKI-Toeslag' for Top Consortia for Knowledge and Innovation (TKI’s) of the Ministry of Economic Affairs and Climate Policy. This work is within the framework of the Institute of Sustainable Process Technology.

\end{acknowledgments}

\clearpage
\appendix
\section{Geometries used for the rheology and rheo-MRI measurements}

\begin{figure}[h!]
\centering
\includegraphics[scale=0.7]{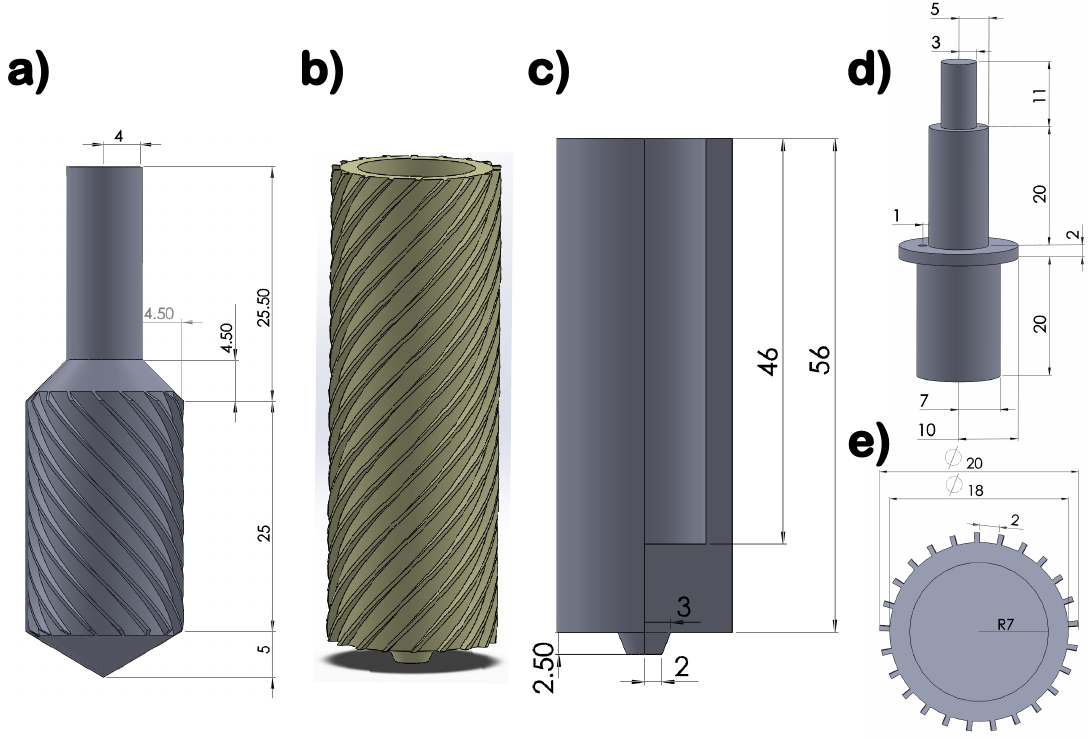}
\caption {\textbf{a)} Front view of the angled CC17 geometry used in the rheological experiments. \textbf{b)} Front view of the geometry used in the rheo-MRI experiments, showing ridges at an angle $\theta$ relative to the vertical axis. \textbf{c)} Front view of the smooth CC20 rheo-MRI geometry (front right is removed to show the hollow inside). \textbf{d)} Cap used to connect the CC20 rheo-MRI geometry to the driving shaft of the rheo-MRI. \textbf{e)} Top view of the angled CC20 rheo-MRI geometry. All dimensions are in mm.}
\label{fig:rheo_rheoMRI_geo_SI}
\end{figure}

\newpage
\section{Flow curves of yield stress fluids}
\label{subsec:flow_curve_YSF}

\begin{figure}[h!]
\centering
\includegraphics[scale=0.7]{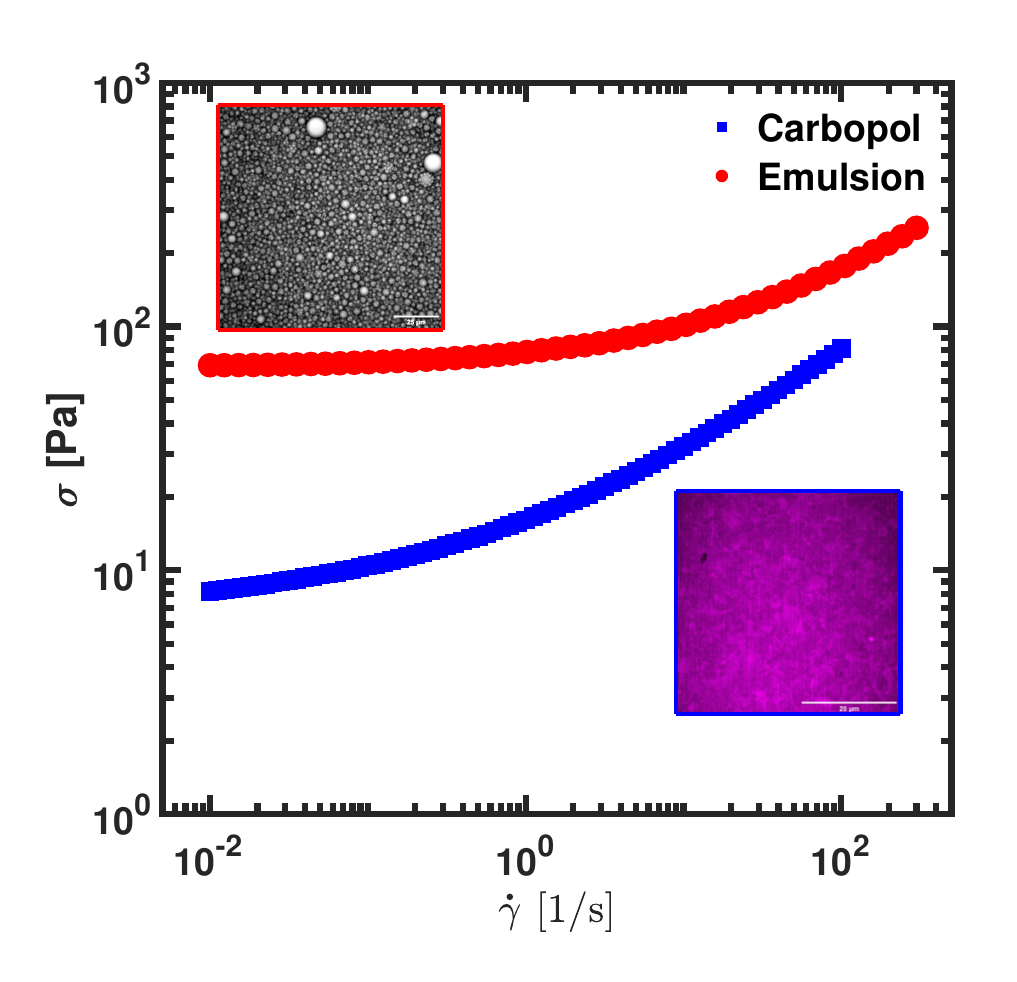}
\caption {Flow curves of Carbopol (\raisebox{.1ex}{\footnotesize\textcolor{blue}{$\blacksquare$}}) and castor oil-in-water emulsion (\textcolor{red}{$\bullet$}), showing Herschel-Bulkley model behaviour. The fitted model parameters for Carbopol: $\sigma_\textnormal{y} = 7.938 \pm 0.696$~Pa,  $k = 8.177 \pm 0.529$~Pa~s$^n$, $n = 0.476 \pm 0.007$ and for castor oil-in-water emulsion: $\sigma_\textnormal{y} = 68.113 \pm 0.066$~Pa,  $k= 10.321 \pm 0.049$~Pa~s$^n$, $n = 0.507 \pm 8.760 \times 10^{-4}$. Insets show the confocal microscopy images of the respective fluids.}
\label{fig:shearrate_ramp_Carbopol_Emulsion_SI}
\end{figure}

We measured the steady state flow curves of Carbopol and castor oil-in-water emulsion using standard geometries in a rheometer, as shown in Figure~\ref{fig:shearrate_ramp_Carbopol_Emulsion_SI}.

\newpage
\section{Rotational and axial velocity profiles for all three fluids in different angled ridged geometries}

For all the cases, we show the axial velocity normalised with the maximum axial velocity within the ridge gap. We observe that in the Newtonian fluid case, all the normalised axial velocity curves collapse onto each other indicating that these velocity profiles only differ in magnitude and the axial velocities are independent of the imposed rotation rate. Please note that for
the 20$^\circ$ ridge angle, the axial velocity curve at highest imposed rotation rate does not collapse onto the rest of the curves, likely due to a measurement error. This curve appears to be an outlier, as it crosses zero at a markedly different position compared to the other curves. 

\begin{figure}[h!]
\centering
\includegraphics[scale=0.28]{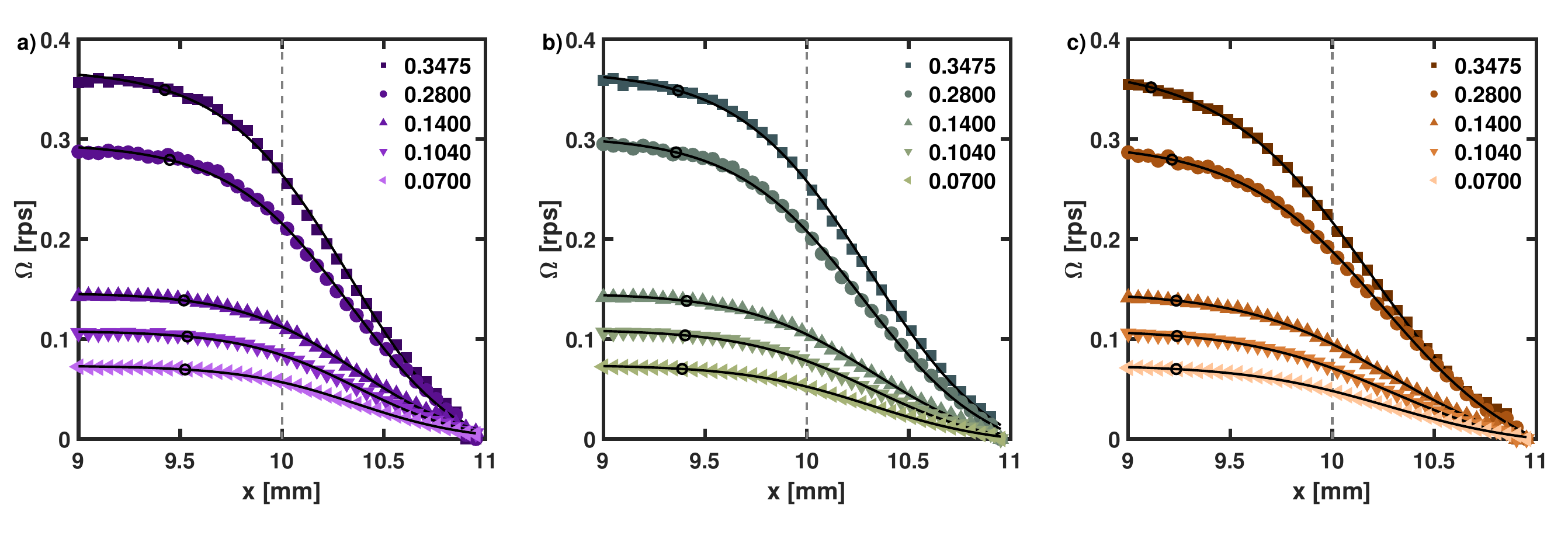}
\includegraphics[scale=0.28]{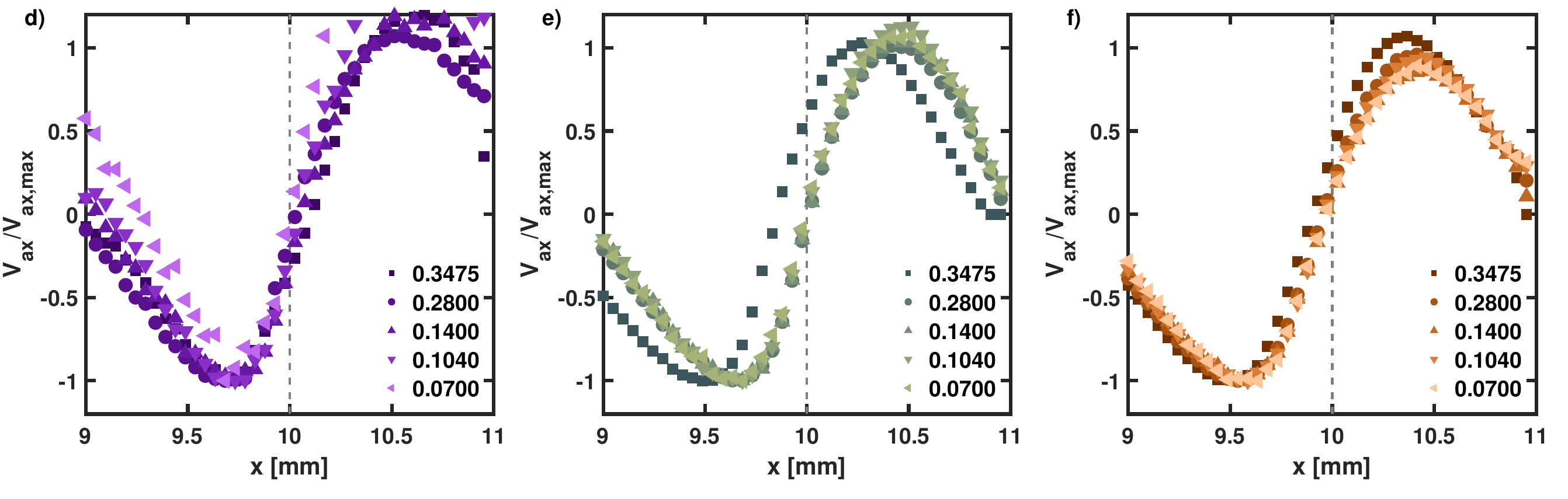}
\caption {Rotational velocity \textbf{(a--c)} and axial velocity \textbf{(d--f)}, normalised with respect to the maximum axial velocity as a function of position in the gap for castor oil, measured at different imposed rotation rates. Each panel presents velocities in geometries with ridges at three different angles: purple for 10$^\circ$, green for 20$^\circ$, and brown for 45$^\circ$. Legends denote the specific imposed rotation rates, $\Omega_\textnormal{o}$ in rps, for each curve, while the dashed line marks the tip of the ridge. The solid lines are fits to a sigmoidal function (Equation~(\ref{ch3:eq:sigmoidalfunction})), from which the penetration depths are obtained. The black circles indicate the position in the gap where each curve decreases by 5$\%$ of the amplitude, defined as the difference between the plateau and minimum values of the curve.}
\label{fig:velprof_Newtonian_SI}
\end{figure}

For the two yield stress fluids, the normalised axial velocity profiles show significant scatter, particularly in the 10$^\circ$ ridge angle case because we divide the axial velocities by their respective small maximum axial velocity value.

\begin{figure}[h!]
\centering
\includegraphics[scale=0.28]{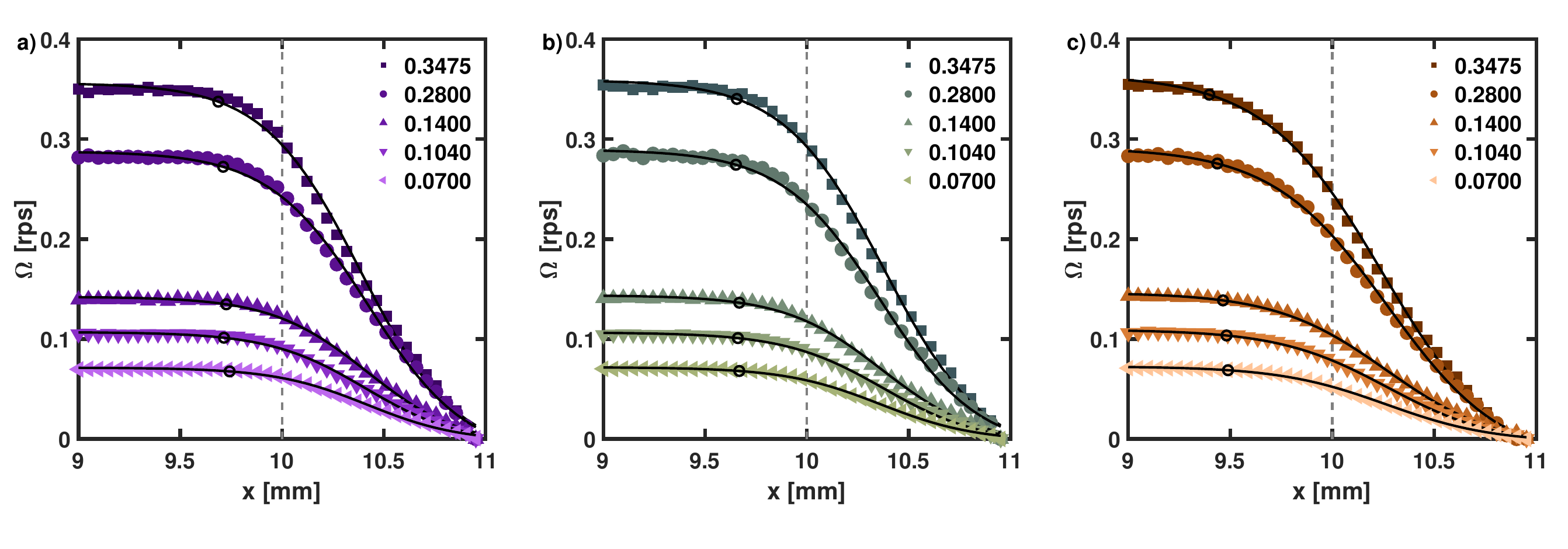}
\includegraphics[scale=0.28]{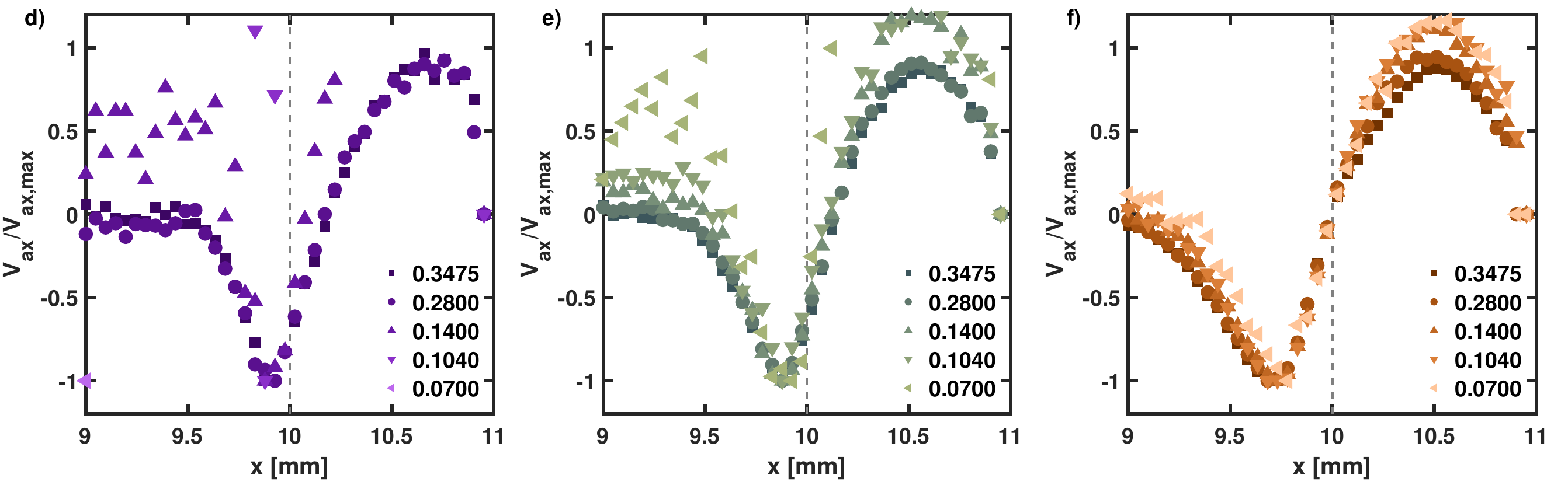}
\caption {Rotational velocity \textbf{(a--c)} and axial velocity \textbf{(d--f)}, normalised with respect to the maximum axial velocity as a function of position in the gap for Carbopol, measured at different imposed rotation rates. Each panel presents velocities in geometries with ridges at three different angles: purple for 10$^\circ$, green for 20$^\circ$, and brown for 45$^\circ$. Legends denote the specific imposed rotation rates, $\Omega_\textnormal{o}$ in rps, for each curve, while the dashed line marks the tip of the ridge. The solid lines are fits to a sigmoidal function (Equation~(\ref{ch3:eq:sigmoidalfunction})), from which the penetration depths are obtained. The black circles indicate the position in the gap where each curve decreases by 5$\%$ of the amplitude, defined as the difference between the plateau and minimum values of the curve.}
\label{fig:velprof_Carbopol_SI}
\end{figure}
\begin{figure}[h!]
\centering
\includegraphics[scale=0.28]{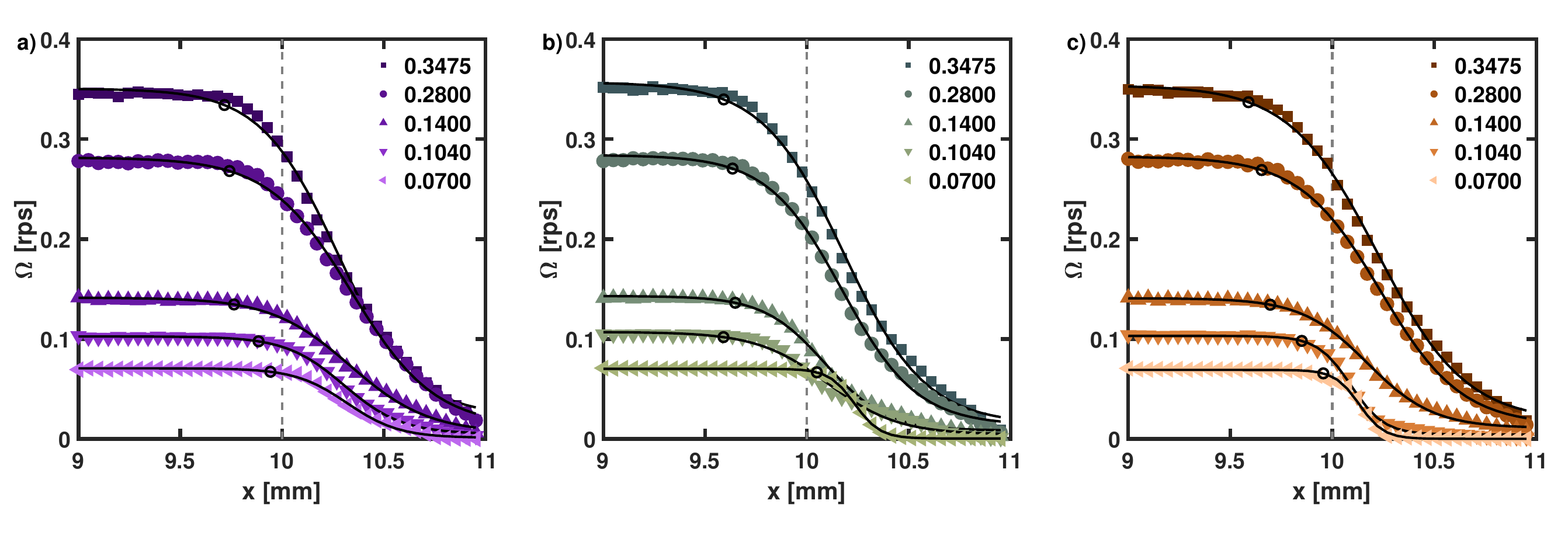}
\includegraphics[scale=0.28]{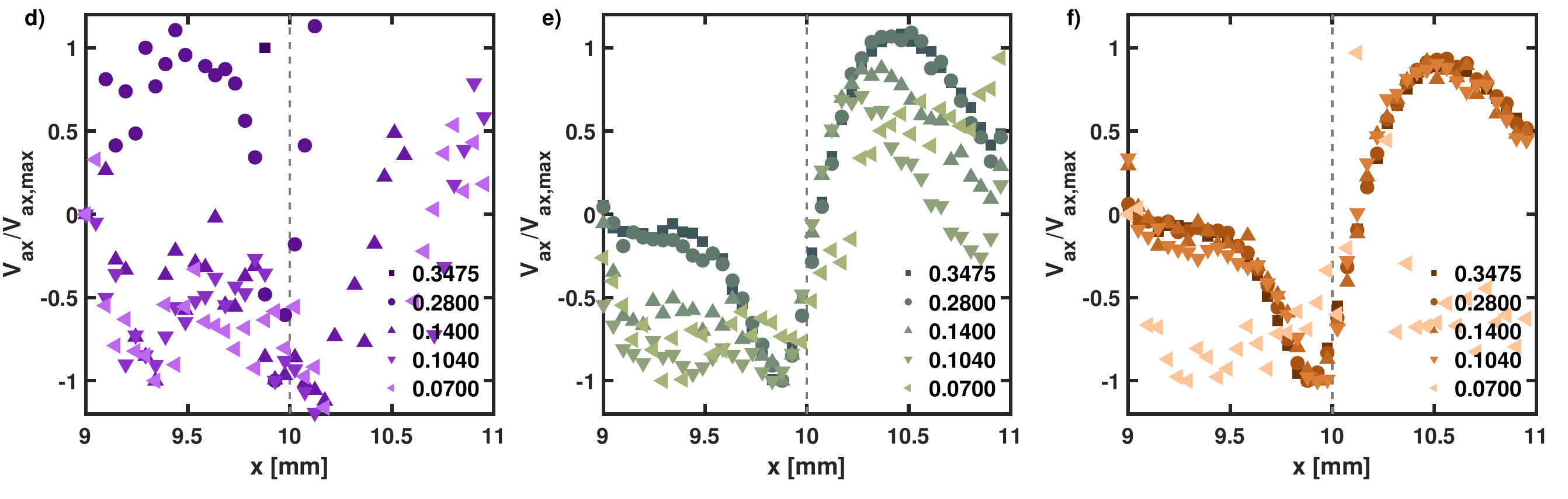}
\caption {Rotational velocity \textbf{(a--c)} and axial velocity \textbf{(d--f)}, normalised with respect to the maximum axial velocity as a function of position in the gap for castor oil-in-water emulsion, measured at different imposed rotation rates. Each panel presents velocities in geometries with ridges at three different angles: purple for 10$^\circ$, green for 20$^\circ$, and brown for 45$^\circ$. Legends denote the specific imposed rotation rates, $\Omega_\textnormal{o}$ in rps, for each curve, while the dashed line marks the tip of the ridge. The solid lines are fits to a sigmoidal function (Equation~(\ref{ch3:eq:sigmoidalfunction})), from which the penetration depths are obtained. The black circles indicate the position in the gap where each curve decreases by 5$\%$ of the amplitude, defined as the difference between the plateau and minimum values of the curve.}
\label{fig:velprof_Emulsion_SI}
\end{figure}
%

\bibliography{References3}

\end{document}